\documentclass[12pt,preprint]{aastex}

\shorttitle{Interstellar HI profiles}
\title{Gaussian Decomposition of $\lambda$21-cm Interstellar HI profiles}

\author{
G. L. Verschuur\altaffilmark{1},
J.T. Schmelz\altaffilmark{2}
}
\altaffiltext{1}{verschuur@aol.com}
\altaffiltext{2}{USRA, 7178 Columbia Gateway Drive, Columbia, MD 21046, jschmelz@usra.edu}

\usepackage{hhline}

\begin{document}

\slugcomment{For submission to Ap.J. April 2020}


\begin{abstract}
Following an established protocol of science $-$ that results must be reproducible $-$ we examine the Gaussian fits to Galactic $\lambda$21-cm emission profiles obtained by two seemingly complementary methods: the semi-automated approach based on the method used by Verschuur (2004) and the automated technique of Nidever et al. (2008). Both methods use data from the Leiden/Argentine/Bonn all-sky survey. The appeal of an automated routine is great, if for no other reason than the time saved over semi-automated fits. The pitfalls, however, are often unanticipated, and the most important aspect of any algorithm is the reproducibility of the results. The comparisons led to the identification of four problems with the Nidever et al. (2008) analysis: (1) different methods of calculating the $\tilde{\chi}^{2}$ measuring the goodness of fit; (2) an ultra-broad component found bridging the gap between low- and intermediate-velocity gas; (3) the lack of an imposed spatial coherence allowing different components to appear and disappear in profiles separated by a fraction of a beam width; and (4) multiple, fundamentally different solutions for the profiles at both the North and South Galactic Poles. A two-step method would improve the algorithm, where an automated fit is followed by a quality-assurance, visual inspection. Confirming evidence emerges from this study of a pervasive component with a line width of order 34 km s$^{-1}$, which may be explained by the Critical Ionization Velocity (CIV) of helium. Since the Nidever et al. (2008) paper contains the only result in the refereed literature that contradicts the CIV model, it is important to understand the flaws in the analysis that let to this contradiction.
\end{abstract}

\section{Introduction}

A series of large-area $\lambda$21-cm emission surveys of Galactic neutral hydrogen from various telescopes are now publicly available. These include the Leiden/Argentine/Bonn (LAB) all-sky survey (Kalberla et al. 2005), the Parks Galactic All Sky Survey (GASS; McClure-Griffiths et al. 2009), the Galactic Arecibo L-Band Feed Array (GALFA) HI survey (Peek et al. 2011), the Effelsberg-Bonn HI Survey (EBHIS; Winkel et al. 2016), and the Green Bank Telescope HI Intermediate Galactic Latitude Survey (GHIGLS; Martin et al. 2015). These high-resolution, high-sensitivity, high-dynamic-range observations are relatively free of side lobe effects, and have sparked various investigations into the nature of the interstellar medium. These analyses reveal that the HI is in the form of complex, tangled filaments (see, e.g., Clark et al. 2014; Martin et al. 2015; Kalberla et al. 2016; Verschuur et al. 2018) that appear to follow the magnetic field. These results inspire new questions of exactly how the interstellar neutral hydrogen is tied to the magnetic field.

Verschuur \& Peratt (1999) and Verschuur \& Schmelz (2010) investigated an interesting manifestation of this magnetic field$-$interstellar hydrogen relationship. They found that Gaussian analysis of interstellar hydrogen by multiple authors studied over many decades using data from different telescopes reveals a pervasive 34 km s$^{-1}$ wide component. The traditional explanation, that the line width results from a kinetic temperature, would mean that T = 24,000 K, high enough to ionize the gas so it could not contribute to the 21-cm profile. Turbulent motions could explain a pervasive broad component, but not why it always has the same numerical value, e.g., 33.7$\pm$ 2.4 km s$^{-1}$ from the literature summary by Verschuur \& Schmelz (2010). Confusion due to telescope side lobes has been proposed as a possible explanation, but the broad feature persists in side-lobe-corrected survey data. They suggested that the 34 km s$^{-1}$ component might result from a well-studied plasma phenomenon called the Critical Ionization Velocity (CIV), where atoms become ionized in the presence of a magnetic field when their kinetic energy relative to the plasma and normal to the magnetic field is equivalent to the ionization potential. The CIV for helium is 34.3 km s$^{-1}$, which could account for the pervasiveness of this 21-cm component.  Peratt \& Verschuur (2000) attempted to account for this phenomenon and further research on the issue will benefit from the stimulus that more and better data would provide. 

These efforts should have been supported by the results from a succession of new, automated analysis programs that fit 21-cm HI line profiles from these large surveys with a series of Gaussian components (e.g., Haud 2000; Nidever et al. 2008; Marchal et al. 2019). These papers describe the pitfalls of Gaussian analysis - velocity blending, ambiguities of the number of components, non-Gaussian profiles, noise peaks, plurality of solutions. To this list we must add the insidious effect of low-level interference signals that will distort the results in unpredictable ways.  Marchal et al. (2019) describe the main opposition to Gaussian decomposition of emission spectra -- that any spectrum can be decomposed if enough Gaussians are used. With all of these drawbacks, how can one be sure that this method provides any real physical information about the emitting gas? 

When describing their distribution of Gaussian parameters for low-velocity gas in their Figure 3b, Nidever et al. (2008) find that their distribution of line widths is not sharply peaked as, e.g., those in Verschuur \& Schmelz (2010). Rather, it has a long tail revealing a continuum of very broad line components. To our knowledge, this plot is the only result in the refereed literature that is in stark contrast with the aforementioned claims that the 34 km/s wide feature is pervasive.

Since Nidever et al. (2008) and Verschuur \& Schmelz (2010) used data from the LAB survey, one wonders why they did not get the same results. If the Gaussian analysis does indeed produce physically meaningful solutions, one would hope that a semi-automated, detailed inspection of hundreds of profiles would produce the same Gaussian parameters as an objective, automated analysis of the approximately 260,000 HI profiles in the LAB survey at 0.\arcdeg5 intervals in both Galactic longitude, {\it l} and latitude, {\it b}.

It is a great step forward for astronomy that surveys like LAB, GASS, GALFA, EBHIS, and GHIGLS are now commonly available to all interested researchers. This is fast becoming the norm rather than the exception. What is not common, however (at least not yet), is for the Gaussian analysis done using these data to be made publicly available. This would allow a subset of the profiles to be inspected individually, perhaps as part of a summer-student, intern, or citizen-science project, to ensure the validity of the automated analysis and to be used for other investigations.

In this paper, we take advantage of the fact that Nidever kindly made his Gaussian fits available to one of us (GLV) for a project unrelated to this analysis. This allowed us to investigate the data in detail and to compare the semi-automated approach based on the work of Verschuur (2004) to the automated fits described by Nidever et al. (2008) for the same LAB data with the goal of understanding the nature of the 34 km s$^{-1}$ wide Gaussian component revealed by both methods. 

\section{Analysis} 

The LAB\footnote[1]{https://www.astro.uni-bonn.de} survey of $\lambda$21-cm emission of Galactic HI results from merging the Leiden/Dwingeloo survey (Hartmann \& Burton 1997) with the Instituto Argentino de Radioastronomia survey (Arnal et al. 2000; Bajaja et al. 2005). The data were corrected for stray radiation at the University of Bonn, with residual errors in the profile wings of less than 20-40 mK. The rms noise is 0.07-0.09 K, and the angular resolution is $~$0.6\arcdeg. The velocity coverage is $-$450 $<$ v$_{LSR}$ $<$ $+$400 km s$^{-1}$, with a resolution of 1.3 km s$^{-1}$.

Gaussian decomposition provides information about the physics underlying the production of the HI profiles in any given direction. For example, the line widths themselves are traditionally assumed to reveal the temperature of the gas, and the area of a Gaussian shape depends on the HI column density for that component. 

The semi-automated Gaussian fitting method uses the Microsoft Excel Solver algorithm described by Verschuur (2004). Solver employs the generalized reduced gradient (GRG2) nonlinear optimization code developed by L. Lasdon, University of Texas at Austin, and A. Waren, Cleveland State University.\footnote[2]{References to papers published on this method may be found at http:// www.optimalmethods.com.} Consider, for example, a set of profiles that map a feature. The initialization of Solver is done on the first profile in the series where the Gaussians themselves and the residual produced by this initial attempt are displayed graphically in real time. Solver is then run and the display is regularly updated as the algorithm works toward a solution, defined by a minimum in the residual signal and a value for reduced chi-squared close to unity. That solution is then downloaded and stored. Then the next profile is automatically considered by Solver, using the previous solution as the initial setting. This process is repeated for all profiles in the map. 

Because we can visually inspect the quality of each solution, the existence of possible problems can be recognized before the solution is downloaded. For example, in some cases non-noise signals (low-level interference) may distort the solution. A common case involves small peaks a few channels wide just off the edge of a profile, which can drag a Gaussian component out to a larger line width than dictated by the HI emission profile itself. Or, in some cases, narrow spikes are found within the velocity extent of the HI emission whose main effect is to create higher reduced chi-squared values than may have been found for an adjacent solution. In most cases, unless the interference occurs inside the extent of the HI profile, the signature of a distortion created by interference can be removed and the algorithm rerun. 

Note that the flow chart outlining the semi-automated Gaussian fitting method in Fig. 2 of Henshaw et al. (2016) is similar to the one described above for Solver. The exception is that their $``$write to file$"$ is done automatically where ours is done manually. This gives us the opportunity to monitor the quality of individual solutions in real time.

Verschuur \& Schmelz (2010) tested the Solver algorithm on 150 synthetically generated profiles. Each input profile consisted of three components, 6, 14 and 34 km s$^{-1}$ wide, plus random noise. The center velocities of each of the components were varied randomly within a range of order 5 km s$^{-1}$. After Gaussian analysis, the histogram of all the test profiles showed peaks at 6.4 $\pm$ 1.0, 13.0 $\pm$ 2.4, and 34.4 $\pm$ 2.3 km s$^{-1}$. See their Fig. 10, which displays the effect of noise on the input parameters.

The automated Gaussian analysis program written by Nidever et al. (2008) in the Interactive Data Language (IDL)\footnote[3]{A product of ITT Visual Information Systems, formerly Research Systems, Inc.} uses the least-squares minimization curve-fitting package MPFIT written by C. Markwardt.\footnote[4]{Available at http://cow.physics.wisc.edu/~craigm/idl/idl.html.} 

Both methods are widely used and mathematically sound. One of us (JTS) has used IDL MPFIT-based programs to analyze X-ray and EVU data to determine solar coronal temperature distributions Schmelz et al. (2010) \& Schmelz et al. (2013). Both methods require a first guess for the Gaussian parameters. Verschuur (2004) and Nidever et al. (2008) have both done tests to determine the reproducibility of the results with different initial guesses. Both algorithms begin with one Gaussian component and only add additional components if the residuals from the previous iteration still show significant structure. (Note: This analysis assumes that the 21-cm profiles can indeed be fit with Gaussian components. This seems to be a reasonable assumption when we examine profiles at high Galactic latitudes where the emission is expected to be optically thin. If, however, the emission features were inherently non-Gaussian, significant residuals might result if one fits it with Gaussian functions.) Since the best fit minimizes both the rms of the residual and the number of Gaussians, both authors attempt to remove components that do not significantly improve the fit to the profiles. 

After establishing these fundamental similarities, we set out to compare the results of the two fitting methods. We first targeted simple profiles, considering directions already identified by Verschuur (2004), profiles analyzed as part of an ongoing study, as well as areas near the North and South Galactic Poles.  To this list we added other profiles from a list where Haud (2000) and Verschuur (2004) Gauss fit results were compared, as well as a series of directions at {\it b} $=$ 60\arcdeg\ considered by us in an unrelated investigation.  Taken together, the choice of directions can be regarded as pseudo-random, except that we avoided low latitude profiles or any requiring more than 10 Gaussian components.  The typical number was from 4 to 6 per profile.

\subsection{Broad Underlying Components in Simple Profiles}

Fig. 1a-c shows a series of profiles where brightness temperature is plotted as a function of velocity. The panels are scaled to a maximum temperature of 2.0 K to highlight the details of the fit rather than show the full amplitude of the features. The results of the Gaussian deconvolution are plotted over the profile with the best-fit solution from the Solver algorithm (labeled as $``$Our analysis$"$) on the left and the Nidever et al. (2008) results (labeled as $``$NMB08 analysis$"$) on the right. The Galactic coordinates ({\it l, b}) are shown in the panels as well as values of the reduced chi-squared calculated using a Restricted Range (RR) of channels covering only the extent of the HI profiles, and the Full Range (FR) using all 777 channels of available data (see below for more details on this important distinction). 

Fig. 1a-c all reveal a component with a width of about 34 km s$^{-1}$, which is highlighted in red. Both methods also use the same number of Gaussians to obtain fits with similar residuals, which are shown in blue. There are many examples of this agreement, which no doubt led to the distribution of peaks in, e.g., Fig. 4 of Verschuur \& Peratt (1999), Fig. 3b of Nidever et al. (2008), and Figures 4a and 5a of Verschuur \& Schmelz (2010). But our goal here is to try to understand where and how the results disagree, especially in the context of the CIV model. In other words, why does Figure 3b of Nidever et al. (2008) have a long tail of very broad line components where the distributions shown by Verschuur (2004) and Verschuur \& Schmelz (2010) are more sharply peaked?

We initially and somewhat naively assumed that both fitting algorithms would get similar results for relatively simple profiles. This assumption is born out in Figs. 1a-c, where we obtain nearly identical results, but the next three examples, Figs. 1d-f, reveal a disagreement that we were not expecting. Both fitting algorithms show the 34 km s$^{-1}$ wide line in red, but the Nidever et al. (2008) fit uses fewer Gaussian components. In addition, their best-fit solution shows clear residuals in blue. 

This puzzled us until we realized that we were using two different ways of calculating the $\tilde{\chi}^{2}$ to measure the goodness of fit. Our method includes only the channels defining the extent of the HI profile where the method of Nidever et al. (2008) included the full velocity range. The $\tilde{\chi}^{2}$ entries shown in each panel can help us understand the significance of these different calculations. The first value uses the Restricted Range (RR), typically from about $-$60 to $+$20 km s$^{-1}$. The second used the Full Range (FR) of the available data, from $-$400 to $+$400 km s$^{-1}$.\footnote[5]{The Nidever et al. (2008) error analysis extended to $-$450 km s$^{-1}$, but this additional 50 km s$^{-1}$ is not available from the Bonn web interface. This additional range contributes mostly noise and does not significantly affect the results.}

Please note that although chi-square minimization (and the goodness of fit testing using the reduced chi square value) itself is not without problems, it is still widely used in the astronomy community. A better criterion for model selection seems to be the corrected version of the Akaike information criterion, which aims to give a better balance between reducing the residual emission and the simplicity of the fit results (see Andrae et al. 2010).

In each example in Figs. 1d-f, the RR and FR values are similar for our analysis (left), but the RR is greater (or much greater) than the FR value for the Nidever et al. (2008) analysis (right). The residuals found in the Nidever et al. (2008) analysis also show a clear need for additional components in the range between about 0 \& $-$20 km s$^{-1}$, which our analysis identified.

Nidever et al. (2008) explain in detail their process for removing Gaussians that do not significantly improve the fits to the velocity profiles. Although it is certainly true that the best fit minimizes both $\tilde{\chi}^{2}$ and the number of Gaussians, how $\tilde{\chi}^{2}$ is calculated is vital for this process. Since the main scientific goal of the Nidever et al. (2008) analysis was to explore the Magellanic Stream, which stretches from roughly $-$400 to $+$300 km s$^{-1}$ over a limited area of sky, it makes sense to use the full range to determine the goodness of fit. The problem becomes apparent when this same method is applied to profiles like those in Fig. 1 where the emission extends over a limited velocity range and does not include structure from the Magellanic Stream. Since the baseline fits the noise extremely well over hundreds of km s$^{-1}$, this artificially improves the goodness of fit and opens up a parameter space where the automated IDL algorithm can remove a Gaussian component that really is required for a good fit to the HI profile.

Table 1 summarizes the data displayed in Fig. 1 with the panel designation listed in the first column, Galactic coordinates ({\it l, b}) of each profile in the second column, and the Gaussian parameters obtained by Solver (Our analysis) in the next three columns followed by the IDL algorithm Nidever et al. (2008) (NMB08 ) in the last 3 columns. Each component is defined by the peak brightness temperature, T$_{B}$, the center velocity, V$_{c}$, and the line width, W (full-width at half-maximum). The line width is traditionally assumed to reveal the kinetic temperature of the gas, but we will argue below that this may not always be the case. Quantitatively, for the 34 km s$^{-1}$ family of these components in these six profiles, the average differences in T$_{B}$, V$_{c}$, $W$ are 0.10 $\pm$ 0.09 K, 0.23 $\pm$ 0.59 km s$^{-1}$, and 0.05 $\pm$ 0.05 km s$^{-1}$. 

\subsection{Bridging the Gap Between Low- and Intermediate-Velocity Peaks}

The next set of examples shown in Fig. 2 are more complex. The profiles show low-velocity as well as intermediate-velocity gas, which tends to peak at around -60 to -70 km s$^{-1}$. The relevant Gaussian parameters are summarized in Table 2, which has the same format as Table 1. In Figs. 2a-c both fitting algorithms reveal a component with a width of order 34 km s$^{-1}$ highlighted in red, and this is always associated with the low-velocity gas peak. For the first two of these examples in Figs. 2a-b, both methods obtain fits with similar RR and FR $\tilde{\chi}^{2}$ values. In the lower right panel, however, where the RR $\tilde{\chi}^{2}$ is high, the problem described in the last subsection is illustrated quite dramatically by the oscillations in the residuals associated with the Nidever et al. (2008) fit to the low-velocity gas.

The profiles of Figs. 2d-f are similar, with low-velocity as well as intermediate-velocity gas, but the fits from each algorithm are quite different. The Solver fit on the left shows the 34 km s$^{-1}$ Gaussian in red, but the IDL algorithm on the right prefers an ultra-broad component, which is highlighted in green. This ultra-broad component, with widths greater than (and often significantly greater than) 40 km s$^{-1}$, tends to bridge the gap between the low- and intermediate-velocity features.

A visual examination of the panels on the right shows significant residuals. There are also problems associated with the $\tilde{\chi}^{2}$ calculation that uses the full velocity range, as described above. In addition, these three profiles (Fig. 2d-f) have something in common $-$ they all have an emission bridge between the low- and intermediate-velocity components. This family of profile shapes allows an ultra-broad component as part of the fit where the previous profiles do not. The profiles in Fig. 1 have only low-velocity emission, and are therefore not wide enough to accommodate an ultra-broad component. The profiles in Fig. 2a-c show no significant bridge emission to accommodate an ultra-broad component.

The problems associated with the ultra-broad component are not simply mathematical. How can we explain the physical nature of the gas associated with this feature? The width cannot indicate a temperature, since it would be of order 100,000 K, hot enough to ionize the hydrogen atoms so they could not contribute to the $\lambda$21-cm profile. It cannot be side lobes, since the LAB data has been side lobe corrected. Turbulent motions could perhaps account for an ultra-broad feature, but current models of the interstellar medium are generally interpreted to imply that the low- and intermediate-velocity gas are in physically different locations in space (Bregman 1980; Kuntz \& Danly 1996; Wakker 2001). How then could a single ultra-broad Gaussian feature cross this significant spatial expanse?

\subsection{Profiles Separated by a Fraction of a Beam Width}

The results of the previous subsection led us to another interesting question $-$ how reproducible are the Nidever et al. (2008) results when their IDL-based algorithm derives fits for adjacent directions offset by a fraction of a beam width?

Fig. 3 has a different format than the previous figures. Figs. 3a-c show Nidever et al. (2008) results for three closely spaced directions with the {\it l, b} values and the RR and FR $\tilde{\chi}^{2}$ indicated in each frame. Figs. 3a-c are for positions around {\it l, b} = 90\arcdeg, 71\arcdeg. These pairs are separated by 0.\arcdeg5 in longitude, equivalent to a third of a beam width at these latitudes. Thus, the derived Gaussian parameters for each pair should be similar. Table 3 summarizes the results. In each case, an ultra-broad component (green lines) on the left of around 42 km s$^{-1}$ wide is not present in the neighboring direction on the right. Instead they have been replaced with ultra-broad components with a widths of 65-70 km s$^{-1}$. These major differences are not physically possible given the small separation in direction.

Figs. 3d-f show the results of our analysis for the same sets of directions as Figs. 3a-c with the relevant data included in Table 3. Here, as is to be expected for pairs of closely spaced directions, the Gaussian parameters are nearly identical without any glaring exceptions. The dominant 34 km s$^{-1}$ family of components are again shown in red. 

Fig. 4 is similar to Fig. 3, except for positions around {\it l, b} = 100\arcdeg, -50\arcdeg. At this latitude, an offset 0.\arcdeg5 in longitude is equivalent to half a beam width on the sky. The Nidever et al. (2008) Gaussian parameters in Figs. 4a-c again include ultra-broad components, and the data in Table 4 show little consistency from one member of the pair to the next. Figs. 4d-f show the results from our analysis for the same pairs with the Gaussian parameters also listed in Table 4. Now small differences between the members of each pair can be recognized in the Table and seen in the figures in a manner that is physically consistent with a change of direction of half-a-beam-width, as is evident in the differences in profile shapes seen in these pairs. 

Nidever et al. (2008) write that they initially used the same procedure as Haud (2000) to force coherence of HI structure in neighboring profiles. If a neighboring profile has a better fit, either smaller rms or fewer Gaussians, (note that although not explicitly mentioned in the original paper, spatial consistency with neighboring solutions might require a fit with a higher rms value of the residual and a greater number of Gaussian components), then this fit is used as the initial guess and the fitting algorithm is rerun. Haud (2000)  allowed his algorithm to re-examine other neighboring profiles, essentially introducing a path for iterating to the best fit. Nidever et al. (2008) found that this iteration was too CPU intensive and opted for a modified scheme that forced the program to return to the previous position after refitting a neighboring position. They also write that this extra step did not improve the solutions substantially, but we suspect that this claim may be subject to the same problems that are described in \S2.1 above. 

In fact, P\"oppel et al. (1994), Haud (2000), Martin et al. (2015) and Miville-Desch\'enes et al. (2017) have all implemented methods that use some information about Gaussian fits to neighboring profiles to favor spatially coherent solutions. The recent algorithm presented by Marchal et al. (2019) takes this spatial coherence to the next level. Where these earlier algorithms simply provide an initial guess to the fit, the optimization is not bound to this guess and can converge to a different solution. The Marchal et al. algorithm, on the other hand, fits all the spectra in a data cube at the same time and imposes spatial coherence by adding what they refer to as specific regularization terms to the cost function.

\subsection{Profiles at the Galactic Poles}

During our search for simple profiles to include in Fig. 1, we made an unexpected discovery concerning the Nidever et al. (2008) database $-$ the IDL-based algorithm had been used to fit the profiles at $\pm$ 90\arcdeg at multiple Galactic longitudes. In other words, {\it different} Gauss fits are included for both the North and South Galactic Poles. We can use this to gain further insight into the reliability and rigor of the IDL-based method and the reproducibility of the Nidever et al. (2008) Gaussian deconvolution process. 

Fig. 5a shows the results of our analysis for the two polar profiles and Figs. 5 b-c show two solutions found in Nidever et al. (2008) data base for each of the poles. The left-hand plots are for the North Galactic Pole and the right-hand plots for the South Galactic Pole. Components with line widths in the 34 km s$^{-1}$ regime are again shown in red, and ultra-broad components in green. Table 5 summarizes the Gaussian parameters. 

All the solutions displayed have similar RR \& FR $\tilde{\chi}^{2}$ values and the residuals in Fig. 5 (blue) do not point to obvious problems. The Gaussian parameter data shown in Table 5 show both similarities and significant differences. The two Gaussian solutions from Nidever et al. (2008) for the North Galactic Pole are clearly inconsistent with each other. The same is true for their two solutions for the South Galactic Pole. 

What may have happened in the Nidever et al. (2008) process of handling the full LAB data set is that the IDL-based algorithm analyzed a series of profiles at a constant longitude, using the fit from the previous latitude as a starting point. Individual series then approached the poles with different fits, resulting in multiple, different fits for the same profile at the pole. This circumstance highlights the need for a rigorous spatial coherence requirement as, e.g., implemented in our analysis or coded in the algorithm of Marchal et al. (2019).

\section{Discussion}

Since Nidever et al. (2008) used the same LAB-survey data as, e.g., Verschuur (2004) and Verschuur \& Schmelz (2010), we were left to wonder why the results of their automated fits were fundamentally different than our semi-automated fits. The fact that the Nidever et al. results are available for comparison gave us the opportunity to follow established scientific protocol and investigate this disagreement, in particular the results related to the pervasive 34 km s$^{-1}$ feature, which is seen by many observers of low-velocity HI gas: e.g., Verschuur (2004), Haud \& Kalberla (2007), Nidever et al. (2008) and Verschuur \& Schmelz (2010 )and references therein.

In the course of this analysis, we have examined about 100 of the approximately 260,000 HI LAB profiles. The comparison of this limited sample was enough to reveal four problems with the Nidever et al. (2008) results that do not necessarily prompt us to throw out their entire endeavor, but do cause us to question some of the assumptions and choices that resulted in the widespread ultra-broad components that comprise the long tail of the distribution shown in their Figure 3b. These problems are: (1) different methods of calculating the $\tilde{\chi}^{2}$ to measure the goodness of fit (\S2.1); (2) the ultra-broad components found to bridge the gap between low- and intermediate-velocity gas (\S2.2); (3) the lack of an imposed spatial coherence that allows fundamentally different components to appear and disappear in profiles separated by a fraction of a beam width (\S2.3); and (4) the multiple, fundamentally different Gaussian solutions for both the North and South Galactic Poles (\S2.4). 

The histograms in Fig. 6a-d summarize the results described in \S2.1 for the two different methods of calculating the $\tilde{\chi}^{2}$ to measure the goodness-of-fit for the directions we examined in the course of this analysis. Figs. 6a-b include only the channels defining the extent of the HI profile, the RR values referred to above. The results from our analysis are on the left and the NMB08 IDL-based algorithm are on the right. The contrast between the panels is striking. The $\tilde{\chi}^{2}$ values obtained by Solver cluster between one and two, with an average of 1.2$\pm$0.3. The distribution obtained for the Nidever et al. fits is far broader, with many values above 2.0 and an average of 1.9$\pm$1.9. (A single value, 16, is off the scale to the right.)

Figs. 6c-d shows how these results change if the goodness-of-fit analysis is done for the full velocity range available in the LAB data. Both distributions are tighter, clustering between values of one and two, with averages of 1.1$\pm$0.2 from Solver and 1.2$\pm$0.3 for the IDL fits. The contrast between the distributions of Fig. 6b and those of Fig. 6d illustrates how the IDL-based algorithm of Nidever et al. (2008) was able to get a $\tilde{\chi}^{2}$ value close to one but also leave residuals that are apparent with a simple visual examination of the profile. (Please note: this is not a problem with IDL or MPFIT; rather, the problem resulted when applying a method that appeared to work for the Magellanic Stream analysis to every profile in the LAB database.)

The histograms in Figs. 6e \& f illustrate how the four problems listed above and described in \S2 affect the Gaussian line widths. These plots isolate the low-velocity gas with center velocities between -30 and 30 km s$^{-1}$. The distribution from Solver on the left reveals a clear peak at 34.2 $\pm$ 1.6 km s$^{-1}$. This is consistent with earlier results of Verschuur (2004) and Verschuur \& Schmelz (2010) and references therein. The results from the IDL-based algorithm on the right show a long tail, similar to Fig. 3b of Nidever et al. (2008).

The goal of the in-depth comparisons presented here was to understand why the Nidever et al. (2008) results did not reveal a peak at 34 km s$^{-1}$ like that in Fig. 6e. Rather, their Fig. 3b shows a long tail revealing a continuum of very broad line width components. Our analysis investigated where and how the results disagree, especially in the context of the CIV model. We call the Nidever et al. (2008) results into question, finding flaws in their use of the IDL algorithm and confirming evidence of a pervasive component with a line width of order 34 km s$^{-1}$. Since the Nidever et al. (2008) paper contains the only result in the refereed literature that contradicts our claims of the importance of the 34 km s$^{-1}$ wide components at low velocities, negating their result, as we do here, paves the way for more robust analysis of this component and its likely relationship to the {\it CIV} effect.  Since this pervasive feature cannot be explained with traditional models, it is important to realize that we may be missing some fundamental aspect of the workings of the interstellar medium. 

Recent results from large-area $\lambda$21-cm emission surveys reveal that the neutral hydrogen gas is in the form of complex, tangled filaments (see, e.g., Peek et al. 2011; Kalberla et al. 2016; Verschuur et al. 2018). These features do not resemble the $``$clouds$"$ depicted in the 20th-century refereed literature; they are much more {\it cirrus} than {\it cumulus}. They do not seem to pay as much attention to gravity, but rather, appear to follow the Galactic magnetic field (see, e.g., Clark et al. 2014; 2015; Martin et al. 2015). The implications of these results are transforming our understanding of the interstellar environment to a place where (like the interplanetary medium in our own solar system) magneto-hydrodynamics may govern and plasma phenomena like the CIV effect may dominate. 

\section{Conclusions}

We attempted to determine the reproducibility of the Nidever et al. (2008) results and understand why some of them were fundamentally different from those of our own analysis $-$ results that led to the hypothesis that the Critical Ionization Velocity effect might be responsible for the pervasive 34 km s$^{-1}$ line width in interstellar HI. This investigation was possible because Nidever had made his Gaussian fits available to one of us (GLV), so his automated fits could be compared directly with our semi-automated fits. During the course of this analysis, we found four fundamental problems with the automated Gaussian analysis of Nidever et al. (2008), which are described in detail in \S2.

1. Values of $\tilde{\chi}^{2}$ derived using the full range of the available multi-channel data may appear to be satisfactory (close to unity), but when the velocity range of the HI emission itself is used, the values are found to be far from ideal. This was revealed in many examples where the residuals emerging from the Gauss fit are seen to be large.

2. The ultra-broad line widths greater than about 40 km s$^{-1}$ are generally found to bridge the gap between peaks at low and intermediate velocities. These components may be mathematically feasible in some cases, but they do not appear to have reasonable physical interpretation.

3. Gauss fitting for profiles separated by less than a beam width reveal changes in the parameters that are inconsistent with such small angular separations on the sky. This comparison is possible because the LAB survey oversampled the sky at very high Galactic latitudes.

4. The presence of multiple solutions at both the North and South Galactic Poles in the Nidever et al. (2008) database seems to result from the lack of an imposed spatial coherence. Each pole direction should have a single solution.

These findings cast doubt on the ultra-broad line widths derived by Nidever et al. (2008). We already know that these line widths cannot be thermal because that would imply kinetic temperatures of order 100,000 K, hot enough to ionize the hydrogen atoms so they could not contribute to the $\lambda$21-cm profile. The ultra-broad line widths cannot be attributed to side lobes, since the LAB data has been side-lobe corrected. If these line widths were produced by turbulent motions in pockets of gas along the full line-of-sight, it would imply that the turbulent signature in one direction is not found in an adjacent direction only a fraction of a beam width away. These results cause us to question the validity of the ultra-broad line widths found in the Nidever et al. (2008) analysis.

The appeal of the Nidever et al. (2008) automated Gauss-fitting routine is great, if for no other reason than the time saved over semi-automated fits. The pitfalls, however, are numerous and often unanticipated. The four problems we found, which are listed above and described in detail in \S2, are based on a visual inspection of roughly 100 relatively simple profiles in directions well away from the Galactic plane. A two-step method would improve the Nidever et al. (2008) results where an automated fit is done first over a broad section of sky, and a visual inspection is done as a quality assurance follow-up. This second step may be tedious, but the analysis presented here indicates that it is necessary, at least until the routine becomes more robust. We suggest that this visual inspection and potential refitting could be done as part of a summer-student, intern, or citizen-science project.

We also note that semi-automated (e.g., Henshaw et al. 2016) and fully automated (e.g. Marchal et al. 2019) algorithms are freely available to the community, which potentially already overcome many of the disadvantages of the Nidever et al. (2008) method that we have identified.

Finally, we have found important reasons to question the long tail in the Gaussian line width distribution seen in Fig. 6f  as well as in the original Nidever et al. (2008) Fig. 3b.  Our analysis provides strong support for the 34 km s$^{-1}$ peak seen in Fig. 6e which pertains to the possibility that this feature might indeed be attributed to the Critical Ionization Velocity signature of Helium.

\section{Acknowledgments}
We are grateful to Dr. P. Kalberla for making the LAB survey data available on-line in a user-friendly form.

\begin{deluxetable}{ccccccccc}
\tablecolumns{9}
\tablewidth{0pc}
\tablecaption{Gaussian parameters for Figure 1}
\tablehead{\colhead{Fig.}	&	\colhead{{\it l, b}}	&	\colhead{T$_{B}$}	&	\colhead{V$_{c}$}	&	\colhead{W}  &	\vline & \colhead{T$_{B}$}	&	\colhead{V$_{c}$}	&	\colhead{W}	\\	\#	&	(\arcdeg)	&	(K)	&	(km s$^{-1}$)	&	(km s$^{-1}$)	& \vline &	(K)	&	(km s$^{-1}$)	&	(km s$^{-1}$)	}
\startdata 
{} & {} & {} &  Our analysis & {}  & \vline & {} & NMB08  & {}\\
\hline
1a & 90, 88 &  1.1 & -17.0 & 35.9 & {} \vline & 1.0 & -17.1 & 37.7 \\ 
{} & {}  & 0.3 & -5.7 & 9.7 & {} \vline & 0.3 & -5.7 & 8.5 \\ 
{} & {}  & 0.4 & -15.0 & 4.7 & {}  \vline& 0.4 & -15.0 & 5.2 \\
{} & $\tilde{\chi}^{2}$ & {} & RR 1.6: FR 1.3 & {} & {}  \vline & {} & RR 1.6: FR 1.3 & {}\\ 
\hline 
1b & 315, -88 & 1.1 & -8.2 & 29.4 & {}  \vline& 1.3 & -8.1 & 28.2 \\ 
{} & {} & 2.8 & -8.6 & 15.1 & {}  \vline & 2.6 & -8.8 & 14.8 \\ 
{} & {} & 1.5 & -8.0 & 5.4 & {}  \vline & 1.5 & -8.0 & 5.4 \\ 
{} & $\tilde{\chi}^{2}$ & {} & FR 0.9: FR 0.9 & {} & {}  \vline & {} & RR 0.9: FR  0.9 & {}\\ 
\hline 
1c & 200, -76 & 1.0 & -9.9 & 30.0 & {}  \vline & 0.9 & -9.4 & 31.0 \\ 
{} & {} & 2.3 & -9.6 & 12.5 & {}  \vline & 2.3 & -9.4 & 13.0 \\ 
{} & {} & 3.1 & -7.1 & 4.2 & {}  \vline & 3.0 & -7.1 & 4.2 \\ 
{} & $\tilde{\chi}^{2}$ & {} & RR 1.2: FR 1.1 & {} & {}  \vline & {} &  RR 1.1: FR 1.1 & {}\\
\hline
1d & 270, -89 & 1.3 & -8.9 & 29.5 &{}  \vline & 1.0 & -10.2 & 29.7\\ 
{} & {} & 2.6 & -7.1 & 13.3 & {}  \vline & 2.8 & -6.6 & 14.1\\ 
{} & {} & 2.7 & -4.3 & 4.3 & {}  \vline & 3.5 & -4.6 & 4.3 \\ 
{} & {} & 0.9 & -5.9 & 2.9 & {}  \vline & {} & {} & {} \\ 
{} & $\tilde{\chi}^{2}$ & {} & RR 1.5: FR 1.5 & {} & {} \vline & {} & RR 17.7: FR 2.8 & {}\\ 
\hline 
1e & 91, 88 & 1.1 & -17.4 & 35.1 & {} \vline  & 0.9 & -17.9 & 40.3 \\ 
{} & {} & 0.4 & -15.0 & 4.1 & {} \vline & 0.4 & -11.5 & 19.3 \\ 
{} & {} & 0.2 & -5.1 & 14.0 & {} \vline & {} & {} & {} \\ 
{} & $\tilde{\chi}^{2}$ & {} & RR 1.7:  FR 1.4 & {} & {}  \vline & {} & RR 2.3: FR 1.5 & {}\\ 
\hline 
1f & 160, -66 & 1.8 & -14.1 & 32.6 & {} \vline & 1.7 & -14.2 & 33.7 \\ 
{} & {} & 3.0 & -7.5 & 13.7 & {} \vline & 3.5 & -7.6 & 13.2 \\ 
{} & {} & 0.6 & -3.9 & 3.1 & {} \vline & {} & {} & {} \\ 
{} & {} & 0.9 & -10.1 & 2.8 & {} \vline & {} & {} & {} \\ 
{} & $\tilde{\chi}^{2}$ & {} & RR 1.4: FR 1.3 & {} & {} \vline & {} & RR 4.7: FR 1.6 & {}\\ 
\enddata
\end{deluxetable}
 \clearpage

\begin{deluxetable}{ccccccccc}
\tablecolumns{9}
\tablewidth{0pc}
\tablecaption{Gaussian parameters for Figure 2}
\tablehead{\colhead{Fig.}	&	\colhead{{\it l, b}}	&	\colhead{T$_{B}$}	&	\colhead{V$_{c}$}	&	\colhead{W}  &	\vline & \colhead{T$_{B}$}	&	\colhead{V$_{c}$}	&	\colhead{W}	\\	\#	&	(\arcdeg)	&	(K)	&	(km s$^{-1}$)	&	(km s$^{-1}$)	& \vline &	(K)	&	(km s$^{-1}$)	&	(km s$^{-1}$)	}
\startdata 
{} & {} & {} &  Our analysis & {} & {} \vline & {} & NMB08  & {}\\
\hline
2a & 183, 62.0 & 0.7 & -6.7 & 38.9 & {} \vline & 0.7 & -6.8 & 39.9 \\ 
{} & {} & 0.8 & -13.2 & 13.7 & {} \vline & 0.8 & -13.1 & 13.9 \\ 
{} & {} & 1.2 & -54.2 & 22.2 & {} \vline & 1.2 & -54.3 & 22.3 \\ 
{} & $\tilde{\chi}^{2}$ & {} & RR 1.0: FR 0.9 & {} & {} \vline & {} & RR 1.0: FR 0.9 & {}\\ 
\hline 
2b & 124.5, 48.0 & 0.5 & -0.2 & 35.2 & {} \vline & 0.7 & 0.8 & 30.1 \\ 
{} & {} & 0.6 & 3.4 & 13.0 & {} \vline & 0.4 & 3.8 & 9.4 \\ 
{} & {} & 1.9 & -53.5 & 30.7 & {} \vline & 1.9 & -53.4 & 31.4 \\ 
{} & $\tilde{\chi}^{2}$ & {} & RR 1.1: FR 1.1 & {} & {} \vline & {} & RR 1.1: FR 1.1 & {}\\ 
\hline 
2c & 95, -46.0 & 12.6 & -4.9 & 11.2 & {} \vline & 12.9 & -4.9 & 10.9 \\ 
{} & {} & 1.9 & -6.6 & 4.2 & {} \vline & {} & {} & {} \\ 
{} & {} & 4.5 & -9.6 & 3.9 & {} \vline & 4.9 & -9.1 & 4.9 \\ 
{}  &  {}  & 0.8 & -9.6 & 31.1 & {} \vline & 1.2 & -8.9 & 30.8 \\ 
{} & {} & 1.2 & -62.9 & 22.4 & {} \vline & 1.2 & -63.2 & 22.5\\ 
{} & {} & 1.1 & -68.6 & 5.1 & {} \vline & 1.1 & -68.5 & 4.9 \\ 
{} & $\tilde{\chi}^{2}$ & {} & RR 1.1: FR 1.0 & {} & {} \vline & {} & RR 1.6: FR 1.0 & {}\\ 
\hline 
2d & 106, -49.5 & 6.7 & -5.4 & 11.7 & {} \vline & 7.0 & -5.5 & 12.7 \\ 
{} & {} & 15.3 & -7.4 & 6.5 & {} \vline &15.4 & -7.4 & 6.4 \\ 
{} & {}  & 0.8 & -8.9 & 32.4 & {} \vline & {} & {} & {} \\ 
{} & {} & 12.8 & -9.5 & 3.1 & {} \vline & 12.9 & -9.4 & 3.1 \\ 
{} & {} & 0.2 & -34.0 & 9.0 & {} \vline & 0.5 & -38.4 & 71.9  \\ 
{} & {} & 1.6 & -53.6 & 21.3 & {} \vline & 0.5 & -46.9 & 9.9\\ 
{} & {} & 1.0 & -55.0 & 7.0 & {} \vline & 2.3 & -56.3 & 11.0 \\ 
{} & {} & 0.6 & -59.0 & 5.0 & {} \vline & {} & {} & {} \\ 
{} & $\tilde{\chi}^{2}$ & {} & RR 1.7: FR 1.5 & {} & {} \vline & {} & RR 2.3: FR 1.7 & {}\\ 
\hline 
2e & 106.5, -49.5  & 0.7 & 0.5 & 3.3 & {} \vline & {} & {} & {}  \\ 
{} & {} & 3.5 & -4.8 & 13.8 & {} \vline & 5.9 & -5.4 & 13.3 \\ 
{} & {}  & 16.9 & -7.2 & 7.1 & {} \vline & 14.0 & -7.0 & 6.5 \\ 
{} & {} & 9.9 & -8.4 & 2.9 & {} \vline & 18.5 & -9.2 & 3.7 \\ 
{} & {} & 10.6 & -10.0 & 2.9 & {} \vline & {} & {} & {}  \\ 
{}  &  {} & 0.6 & -11.9 & 34.1 & {} \vline & 0.4 & -31.3 & 62.9 \\ 
{} & {} & 2.1 & -55.1 & 19.5 & {} \vline & 0.6 & -45.7 & 7.1 \\ 
{} & {} & 2.1 & -56.2 & 5.8 & {} \vline & 4.0 & -57.1 & 11.4\\ 
{} & {} & 1.2 & -60.6 & 4.3 & {} \vline & {} & {} & {} \\ 
{} & $\tilde{\chi}^{2}$ & {} & RR 1.3: FR 1.8 & {} & {} \vline & {} & RR 2.6: FR 2.0 & {}\\ 
\hline 
2f & 104.5, -48 & 2.4 & -2.5 & 5.3 & {} \vline & 0.5 & 6.2 & 6.4 \\ 
{} & {} & 8.1 & -4.5 & 3.2 & {} \vline & 5.8 & -4.3 & 2.7 \\ 
{} & {} & 2.9 & -5.2 & 15.1 & {} \vline & 8.7 & -6.1 & 10.9 \\ 
{} & {} & 19.1 & -10.2 & 6.2 & {} \vline & 16.2 & -10.8 & 6.0 \\ 
{} & {} & 2.0 & -13.2 & 3.0 & {} \vline & 0.5 & -41.2 & 60.6 \\ 
{}  &  {}   & 0.4 & -14.5 & 33.8 & {} \vline & {} & {} & {} \\ 
{} & {} & 0.7 & -48.8 & 20.8 & {} \vline & 0.3 & -47.5 & 11.7 \\ 
{} & $\tilde{\chi}^{2}$ & {} & RR 1.7: FR 1.5 & {} & {} \vline & {} & RR 3.7: FR1.8 & {}\\ 
\enddata
\end{deluxetable}
\clearpage

\begin{deluxetable}{ccccccccc}
\tablecolumns{9}
\tablewidth{0pc}�																			
\tablecaption{Gaussian	parameters	for	Figure	3}�																					\tablehead{\colhead{Fig.}	&	\colhead{{\it l, b}}	&	\colhead{T$_{B}$}	&	\colhead{V$_{c}$}	&	\colhead{W}	&	\colhead{{\it l, b}}	&	\colhead{T$_{B}$}	&	\colhead{V$_{c}$}	&	\colhead{W}	\\	\#	&	(\arcdeg)	&	(K)	&	(km s$^{-1}$)	&	(km s$^{-1}$)	&	(\arcdeg)	&	(K)	&	(km s$^{-1}$)	&	(km s$^{-1}$)		}
\startdata�																						
{}	&	{}	&	{}	&�	NMB08	&	{}	&	{}	&	{}	&	NMB08 	&	{}	\\
\hline
3a	&	90.0, 71.5	&	1.1	&	1.7	&	8.1	&	89.5, 71.5	&	1.3	&	1.3	&	10.8	\\
{}	&	{}	&	0.9	&	-15.5	&	42.7	&	{}	&	0.3	&	-19.1	&	69.6	\\
{}	&	{}	&	0.7	&	-23.8	&	13.2	&	{}	&	1.2	&	-21.9	&	20.5	\\															
{} &   $\tilde{\chi}^{2}$ & {} & RR 1.2: FR 1.1 & {} & {} & {} & RR 1.4: FR 1.1 & {}\\ 					
\hline													
3b	&	90.5, 71.0	&	1.1	&	-24.2	&	16.0	&	90.0, 71.0	&	1.5	&	-22.9	&	17.9	\\
{}	&	{}	&	0.5	&	-18.0	&	43.8	&	{}	&	0.3	&	-19.9	&	65.2	\\
{}	&	{}	&	1.3	&	1.7	&	10.7	&	{}	&	1.4	&	1.3	&	11.5	\\															
{} & $\tilde{\chi}^{2}$ & {} & RR 1.6: FR 1.2 & {} & {} & {} & RR 1.6: FR 1.1 & {}\\ 						
\hline													
3c	&	90.5, 70.5	&	0.7	&	-26.1	&	16.9	&	90, 70.5	&	0.2	&	-21.2	&	69.5	\\									
{}	&	{}	&	0.7	&	-17.8	&	11.3	&	{}	&	1.3	&	-23.1	&	19.4	\\
{}	&	{}	&	1.3	&	2.2	&	42.7	&	{}	&	1.4	&	1.5	&	13.0	\\
{}  & $\tilde{\chi}^{2}$ & {} & RR 0.8: FR 0.8 & {} & {} & {} & RR 0.8: FR 0.8 & {}\\ 					
\hline													
\hline													
{}	&	{}	&	{}	&�	Our analysis	&	{}	&	{}	&	{}	&	Our analysis	&	{}	\\	
\hline																								
3d	&	90.0, 71.5	&	0.8	&�	-10.0	&	34.0	&	89.5, 71.5	&	0.8	&	-10.0	&	34.0	\\
{}	&	{}	&	0.2	&�	2.9	&	14.0	&	{}	&	0.2	&	2.9	&	14.0	\\
{}	&	{}	&	0.8	&�	1.6	&	6.3	&	{}	&	0.8	&	1.6	&	6.3	\\
{}	&	{}	&	0.6	&�	-26.9	&	20.9	&	{}	&	0.6	&	-26.9	&	20.9	\\
{}	&	{}	&	0.5	&�	-23.1	&	8.1	&	{}	&	0.5	&	-23.1	&	8.1	\\
{}	&	{}	&	0.1	&�	-55.0	&	10.0	&	{}	&	0.1	&	-55.0	&	10.0	\\															
{}  & $\tilde{\chi}^{2}$ & {} & RR 1.2: FR 1.1 & {} & {} & {} & RR 1.1: FR 1.1 & {}\\ 					
\hline													
3e	&	90.5, 71.0	&	0.4	&�	-15.7	&	34.4	&	90.0, 71.0	&	0.6	&	-10.6	&	36.7	\\
{}	&	{}	&	1.3	&�	1.8	&	11.5	&	{}	&	1.2	&	1.7	&	10.0	\\
{}	&	{}	&	1.3	&�	-24.5	&	17.6	&	{}	&	1.0	&	-25.0	&	18.3	\\
{}	&	{}	&	{}	&�	{}	&	{}	&	{}	&	0.4	&	-22.2	&	9.7	\\															
{}  & $\tilde{\chi}^{2}$ & {} & RR 1.6: FR 1.2 & {} & {} & {} & RR 1.0: FR 1.1 & {}\\ 					
\hline												
3f	&	90.5, 70.5	&	0.6	&�	-16.9	&	34.6	&    90.0, 70.5	&	0.6	&	-15.6	&	39.6	\\
{}	&	{}	&	1.3	&�	2.4	&	11.8	&	{}	&	1.2	&	1.9	&	11.4	\\		
{}	&	{}	&	0.8	&�	-26.4	&	17.8	&	{}	&	0.9	&	-24.4	&	15.7	\\															
{}  & $\tilde{\chi}^{2}$ & {} & RR 0.9: FR 0.8 & {} & {} & {} & RR 0.8: FR 0.8 & {}\\ 						
\enddata�																					
\end{deluxetable}�																				
\clearpage�

\begin{deluxetable}{ccccccccc}�																																																					
\tablecolumns{9}�																																																					
\tablewidth{0pc}�																																																					
\tablecaption{Gaussian parameters for Figure 4}																																																					
\tablehead{\colhead{Fig.}	&	\colhead{{\it l, b}}	&	\colhead{T$_{B}$}	&	\colhead{V$_{c}$}	&	\colhead{W}	&	\colhead{{\it l, b}}	&	\colhead{T$_{B}$}	&	\colhead{V$_{c}$}	&	\colhead{W}	\\	\#	&	(\arcdeg)	&	(K)	&	(km s$^{-1}$)	&	(km s$^{-1}$)	&	(\arcdeg)	&	(K)	&	(km s$^{-1}$)	&	(km s$^{-1}$)		}
\startdata�													
{}	&	{}	&	{}	&�	NMB08	&	{}	&	{}	&	{}	&	NMB08 	&	{}	\\
\hline														
4a	&	100.5, -50.5	&	11.4	&	-1.7	&	8.1	&	100.0, -50.5	&	11.0	&	-2.7	&	10.4	\\																																				
{}	&	{}	&	1.0	&	-7.9	&	32.4	&	{}	&	1.0	&	-7.6	&	22.0	\\																																				
{}	&	{}	&	5.3	&	-8.9	&	2.5	&	{}	&	6.6	&	-9.0	&	2.8	\\																																				
{}	&	{}	&	18.6	&	-9.3	&	6.1	&	{}	&	14.3	&	-9.2	&	5.8	\\																																				
{}	&	{}	&	0.2	&	-42.5	&	13.4	&	{}	&	0.2	&	-36.9	&	86.1	\\																																				
{}	&	{}	&	0.6	&	-60.9	&	3.7	&	{}	&	{}	&	{}	&	{}	\\																																				
{}	&	{}	&	1.5	&	-63.0	&	20.2	&	{}	&	1.5	&	-63.0	&	15.8	\\																																				
{}	&	{}	&	{}	&	RR 1.0: FR 0.8	&	{}	&	{}	&	{}	&	RR 2.9: FR 1.3	&	{}	\\																																				
\hline																																																					
4b	&	100.5, -51.0	&	8.2	&	-0.1	&	5.9	&	100.0, -51.0	&	10.0	&	0.0	&	5.9	\\																																				
{}	&	{}	&	2.7	&	-5.4	&	19.5	&	{}	&	2.3	&	-5.9	&	21.8	\\																																				
{}	&	{}	&	12.7	&	-7.8	&	8.3	&	{}	&	13.0	&	-7.3	&	8.1	\\																																				
{}	&	{}	&	7.2	&	-10.3	&	4.3	&	{}	&	12.8	&	-10.4	&	4.5	\\																																				
{}	&	{}	&	0.3	&	-41.2	&	64.7	&	{}	&	0.3	&	-41.2	&	47.1	\\																																				
{}	&	{}	&	0.8	&	-64.3	&	18.8	&	{}	&	0.6	&	-64.1	&	16.0	\\																																				
{}	&	{}	&	0.4	&	-61.2	&	3.2	&	{}	&	{}	&	{}	&	{}	\\																																				
{}	&	{}	&	{}	&	RR 1.1: FR1.1	&	{}	&	{}	&	{}	&	RR 0.9: FR 0.7	&	{}	\\																																				
\hline																																																					
4c	&	100.5, -51.5	&	6.0	&	-0.7	&	7.2	&	100.0, -51.5	&	10.2	&	-1.7	&	8.4	\\																																				
{}	&	{}	&	16.1	&	-8.0	&	8.7	&	{}	&	13.2	&	-8.0	&	6.7	\\																																				
{}	&	{}	&	2.4	&	-5.2	&	17.2	&	{}	&	5.0	&	-10.8	&	8.4	\\																																				
{}	&	{}	&	2.0	&	-11.8	&	3.1	&	{}	&	3.0	&	-11.6	&	3.1	\\																																				
{}	&	{}	&	0.4	&	-25.5	&	52.0	&	{}	&	0.5	&	-14.7	&	45.4	\\																																				
{}	&	{}	&	0.6	&	-64.3	&	18.8	&	{}	&	0.2	&	-58.0	&	24.5	\\																																				
{}	&	{}	&	{}	&	RR 1.1: FR1.1	&	{}	&	{}	&	{}	&	RR 1.1: FR 1.0	&	{}	\\																																				
\hline																																																					
\hline																																																					
{}	&	{}	&	{}	&�	Our analysis	&	{}	&	{}	&	{}	&	Our analysis	&	{}	\\
\hline																																				
4d	&	100.5, -50.5	&	0.4	&�	-14.0	&	35.4	&	100.0, -50.5	&	0.5	&	-13.8	&	33.7	\\																																				
{}	&	{}	&	3.2	&�	-4.4	&	16.1	&	{}	&	3.3	&	-3.5	&	16.1	\\																																				
{}	&	{}	&	3.5	&�	0.4	&	6.2	&	{}	&	4.4	&	1.4	&	4.7	\\																																				
{}	&	{}	&	18.4	&�	-9.2	&	5.8	&	{}	&	18.8	&	-9.0	&	5.5	\\																																				
{}	&	{}	&	4.3	&�	-8.8	&	2.3	&	{}	&	4.1	&	-9.0	&	2.4	\\																																				
{}	&	{}	&	6.9	&�	-2.5	&	6.6	&	{}	&	8.0	&	-3.2	&	5.4	\\																																				
{}	&	{}	&	0.9	&�	-59.1	&	24.1	&	{}	&	0.9	&	-60.0	&	21.0	\\																																				
{}	&	{}	&	0.6	&�	-68.4	&	9.9	&	{}	&	0.5	&	-69.0	&	8.0	\\																																				
{}	&	{}	&	0.9	&�	-61.0	&	6.4	&	{}	&	0.8	&	-62.5	&	7.0	\\																																				
{}	&	{}	&	{}	&�	RR 0.7: FR 0.8	&	{}	&	{}	&	{}	&	RR 1.1: FR 1.1	&	{}	\\																																				
\hline																																																					
4e	&	100.5, -51.0	&	0.6	&�	-10.2	&	33.4	&	100.0, -51.0	&	0.7	&	-13.4	&	35.1	\\																																				
{}	&	{}	&	3.4	&�	-5.6	&	16.4	&	{}	&	3.6	&	-4.4	&	15.6	\\																																				
{}	&	{}	&	3.4	&�	-5.4	&	4.4	&	{}	&	4.0	&	-3.1	&	5.5	\\																																				
{}	&	{}	&	8.6	&�	-0.4	&	6.1	&	{}	&	8.3	&	0.3	&	5.3	\\																																				
{}	&	{}	&	9.1	&�	-8.8	&	7.1	&	{}	&	14.3	&	-8.6	&	7.0	\\																																				
{}	&	{}	&	8.0	&�	-10.0	&	4.7	&	{}	&	8.6	&	-10.5	&	4.1	\\																																				
{}	&	{}	&	0.2	&�	-38.2	&	14.0	&	{}	&	{}	&	{}	&	{}	\\																																				
{}	&	{}	&	1.0	&�	-63.4	&	23.2	&	{}	&	0.5	&	-59.5	&	24.3	\\																																				
{}	&	{}	&	0.4	&�	-61.3	&	3.3	&	{}	&	0.3	&	-64.8	&	10.6	\\																																				
{}	&	{}	&	{}	&�	RR 1.0: FR1.1	&	{}	&	{}	&		&	RR 1.2: FR 0.8	&	{}	\\																																				
\hline																																																					
4f	&	100.5, -51.5	&	0.6	&�	-10.9	&	33.5	&	100.0, -51.5	&	0.4	&	-10.7	&	33.6	\\																																				
{}	&	{}	&	4.2	&�	-6.0	&	14.8	&	{}	&	4.7	&	-6.8	&	14.3	\\																																				
{}	&	{}	&	6.5	&�	-1.4	&	7.7	&	{}	&	6.5	&	-0.9	&	7.2	\\																							
{}	&	{}	&	3.1	&�	-6.9	&	4.8	&	{}	&	3.7	&	-7.8	&	4.4	\\																																				
{}	&	{}	&	8.8	&�	-7.6	&	7.3	&	{}	&	9.6	&	-7.4	&	7.5	\\																																				
{}	&	{}	&	6.3	&�	-11.3	&	5.2	&	{}	&	6.2	&	-11.4	&	4.2	\\																																				
{}	&	{}	&	0.2	&�	-36.9	&	15.3	&	{}	&	0.2	&	-36.9	&	15.3	\\																																				
{}	&	{}	&	0.7	&�	-63.4	&	21.5	&	{}	&	0.2	&	-62.8	&	21.5	\\																																				
													
{}	&	{}	&	{}	&�	RR 1.8:  FR1.1	&	{}	&	{}	&	{}	&	RR 1.4: FR 1.1	&	{}	\\	
\enddata�											
\end{deluxetable}�													
\clearpage�

\begin{deluxetable}{ccccccccc} 
\tablecolumns{9} 
\tablewidth{0pc} 
\tablecaption{Gaussian parameters for Figure 5} 
\tablehead{\colhead{Fig.}	&	\colhead{{\it l, b}}	&	\colhead{T$_{B}$}	&	\colhead{V$_{c}$}	&	\colhead{W}	&	\colhead{{\it l, b}}	&	\colhead{T$_{B}$}	&	\colhead{V$_{c}$}	&	\colhead{W}	\\	\#	&	(\arcdeg)	&	(K)	&	(km s$^{-1}$)	&	(km s$^{-1}$)	&	(\arcdeg)	&	(K)	&	(km s$^{-1}$)	&	(km s$^{-1}$)		}
\startdata�																																		
{}	&	{}	&	{}	&�	Our analysis	&	{}	&	{}	&	{}	&	Our analysis	&	{}	\\
\hline																	
5a	&	NGP 	&	0.4	&	-10.6	&	33.8	&	SGP	&	1.3	&	-9.9	&	30.2	\\																	
{}	&	{}	&	0.8	&	-6.1	&	12.3	&	{}	&	1.5	&	-5.3	&	14.2	\\																	
{}	&	{}	&	0.7	&	-24.7	&	20.8	&	{}	&	1.1	&	-4.4	&	4.6	\\																	
{}	&	{}	&	0.2	&	-45.6	&	8.6	&	{}	&	0.9	&	-12.7	&	8.8	\\																	
{}	&	{}	&	{}	&	RR 0.9: FR 1.0	&	{}	&	{}	&	{}	&	RR 1.2: FR 1.1	&	{}	\\																	
\hline																																		
\hline																																		
{}	&	{}	&	{}	&	NMB08  	&	{}	&	{}	&	{}	&	NMB08  	&	{}	\\
\hline																	
5b	&	NGP 	&	0.4	&	-21.3	&	49.8	&	SGP	&	1.0	&	-9.9	&	32.3	\\																	
{}	&	270.0, 90.0	&	1.0	&	-6.5	&	13.4	&	45.0, -90.0	&	2.1	&	-7.5	&	16.9	\\																	
{}	&	{}	&	0.6	&	-23.9	&	15.8	&	{}	&	1.1	&	-4.4	&	4.6	\\																	
{}	&	{}	&	7.2	&	-10.3	&	4.3	&	{}	&	0.4	&	-12.6	&	4.8	\\																	
{}	&	{}	&	{}	&	RR 1.3: FR1.0	&	{}	&	{}	&	{}	&	RR 1.3: FR 1.1	&	{}	\\																	
\hline																																		
5c	&	NGP 	&	1.0	&	-18.6	&	35.6	&	SGP	&	1.3	&	-9.4  &	30.3	\\																	
{}	&	30.0, 90.0	&	0.7	&	-5.5	&	9.9	&	30.0, -90.0	&	1.0	&	-1.6	&	9.8	\\																	
{}	&	{}	&	2.4	&	-5.2	&	17.2	&	{}	&	1.7	&	-11.1	&	10.3	\\																	
{}	&	{}	&	2.0	&	-11.8	&	3.1	&	{}	&	1.4	&	-4.6	&	4.9	\\																	
{}	&	{}	&	{}	&	RR 1.1: FR 1.0	&	{}	&	{}	&	{}	&	RR 1.2: FR 1.0	&	{}	\\																			\enddata 
\end{deluxetable} 
\clearpage 

\begin{figure}
\figurenum{1a-c}
\epsscale{1.0}
\plotone{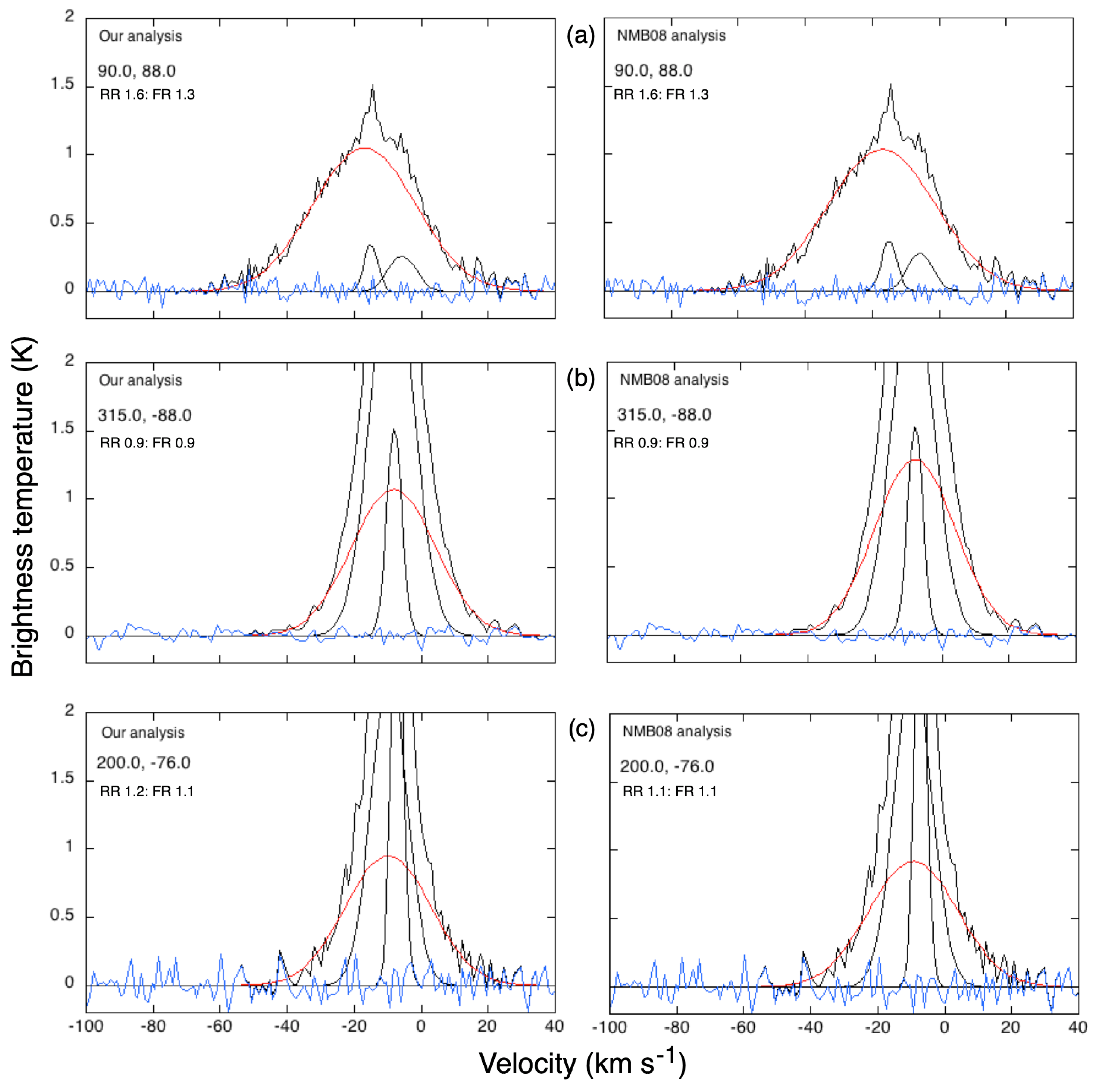}
\caption{ 
Three examples of Gauss fits where the results obtained by the two approaches agree extremely well. The left-hand plots show our fits, and the right-hand plots are the Nidever et al. (2008) results, labeled NMB08. The Gaussians for components with a width of about 34 km s$^{-1}$ are highlighted in red. The residuals after Gauss fitting are shown in blue. The Galactic longitude and latitude coordinates are indicated in each frame. Also shown are the values for $\tilde{\chi}^{2}$ obtained using the Restricted Range (RR) covering only the 60 to 80 channels of data exhibiting Galactic HI emission and the Full Range (FR) that includes all 777 channels of data available in the LAB data base. The parameters for the Gaussian components are listed in Table 1. }
\end{figure}

\clearpage

\begin{figure}
\figurenum{1d-f}
\epsscale{1.0}
\plotone{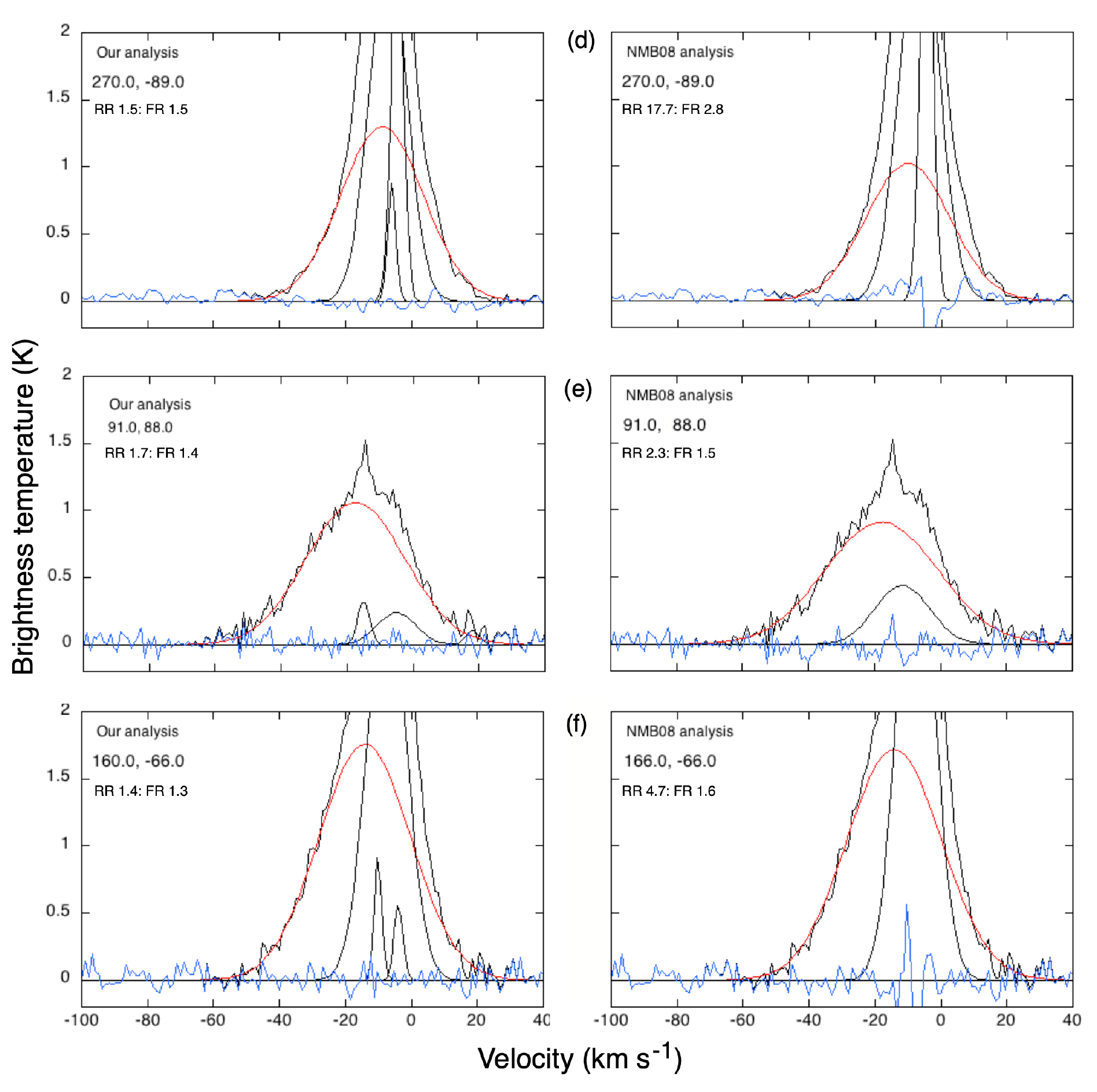}
\caption{Similar to Fig. 1a-c for three directions where we find a good match for the broad component shown in red, but the NMB08 results use fewer Gaussian components and exhibit large residuals (blue). The Gaussian parameters for the respective components are listed in Table 1. 
}
\end{figure}

\clearpage

\begin{figure}
\figurenum{2a-c}
\epsscale{1.0}
\plotone{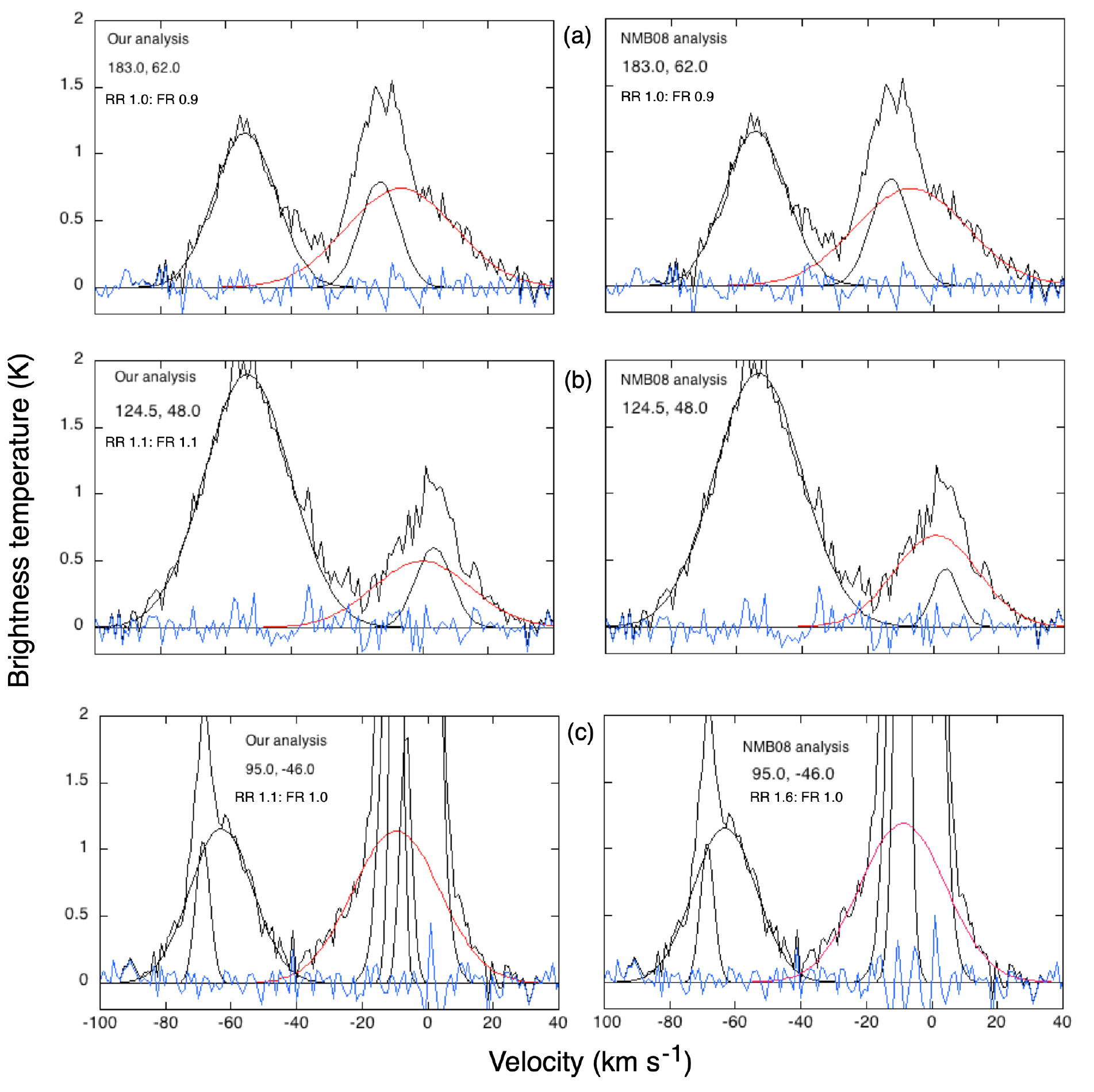}
\caption{Similar to Fig. 1 for three directions where the HI profile has a double-peaked structure that includes intermediate-velocity gas. In these examples, the results of both Gauss fitting methods are similar with broad components shown in red, but note the significant residuals (blue) around zero velocity in the lower right frame.  The spike around $+$2 km s$^{-1}$ may be a weak interference signal that affects the residuals. The Gaussian parameters are summarized in Table 2.
}
\end{figure}

\clearpage

\begin{figure}
\figurenum{2d-f}
\epsscale{1.0}
\plotone{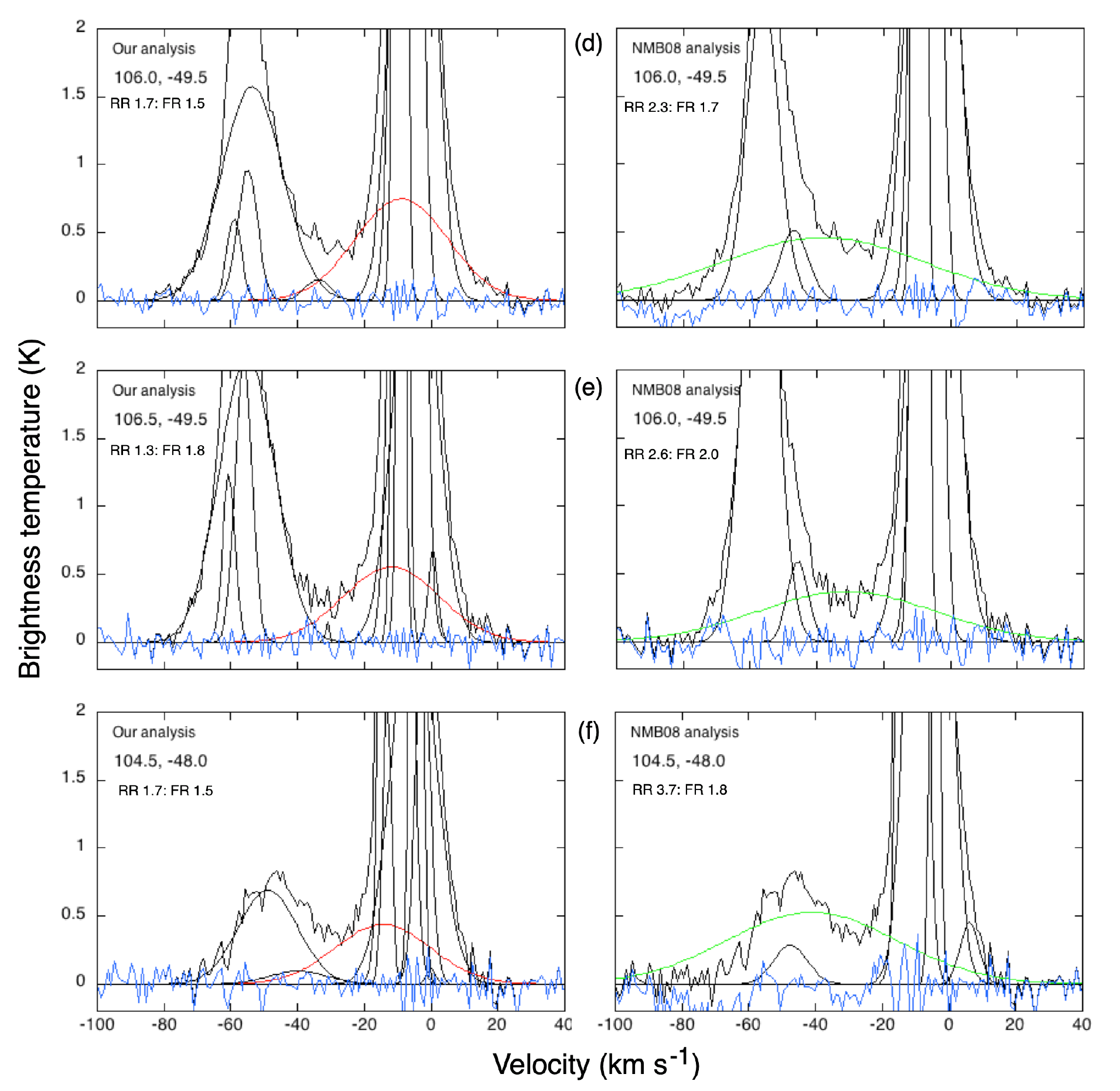}
\caption{ Similar to Fig. 1 for three directions where the NMB08 analysis bridged the gap between low- and intermediate-velocity peaks with ultra-broad component highlighted in green (see text). The Gaussian parameters are summarized in Table 2.
}
\end{figure}

\clearpage

\begin{figure}
\figurenum{3a-c}
\epsscale{0.9}
\plotone{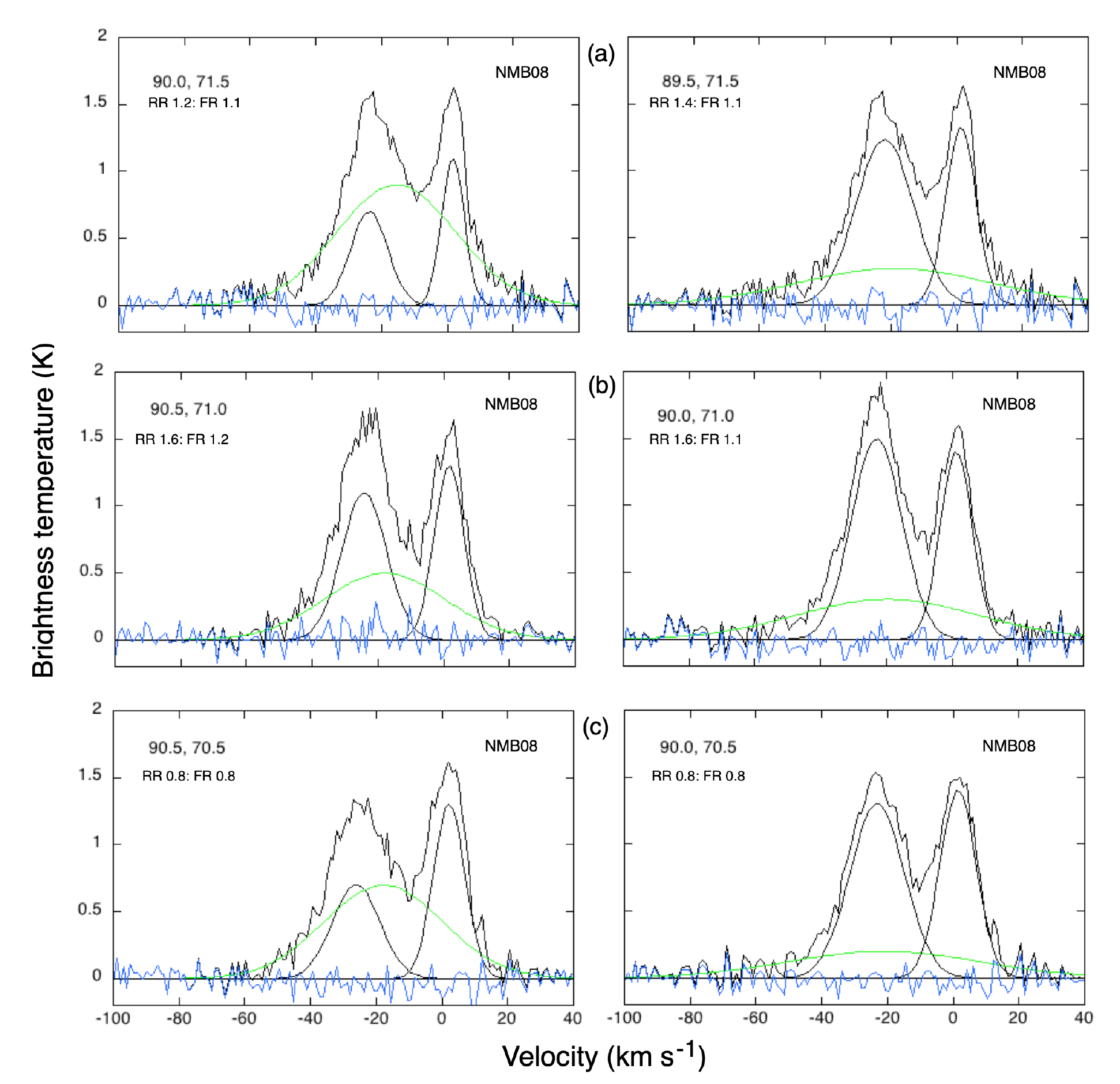}
\caption{Solutions from the NMB08 analysis for closely spaced directions with each pair separated by 0.\arcdeg5 in longitude, which corresponds to a third of a beam width at these latitudes. The Galactic coordinates and the RR \& FR $\tilde{\chi}^{2}$ values are again shown in the frames. Adjacent positions show striking differences for the ultra-broad components, highlighted in green, which are not expected for closely spaced directions a third of a beam width apart. }
\end{figure}
\clearpage

\begin{figure}
\figurenum{3d-f}
\epsscale{0.9}
\plotone{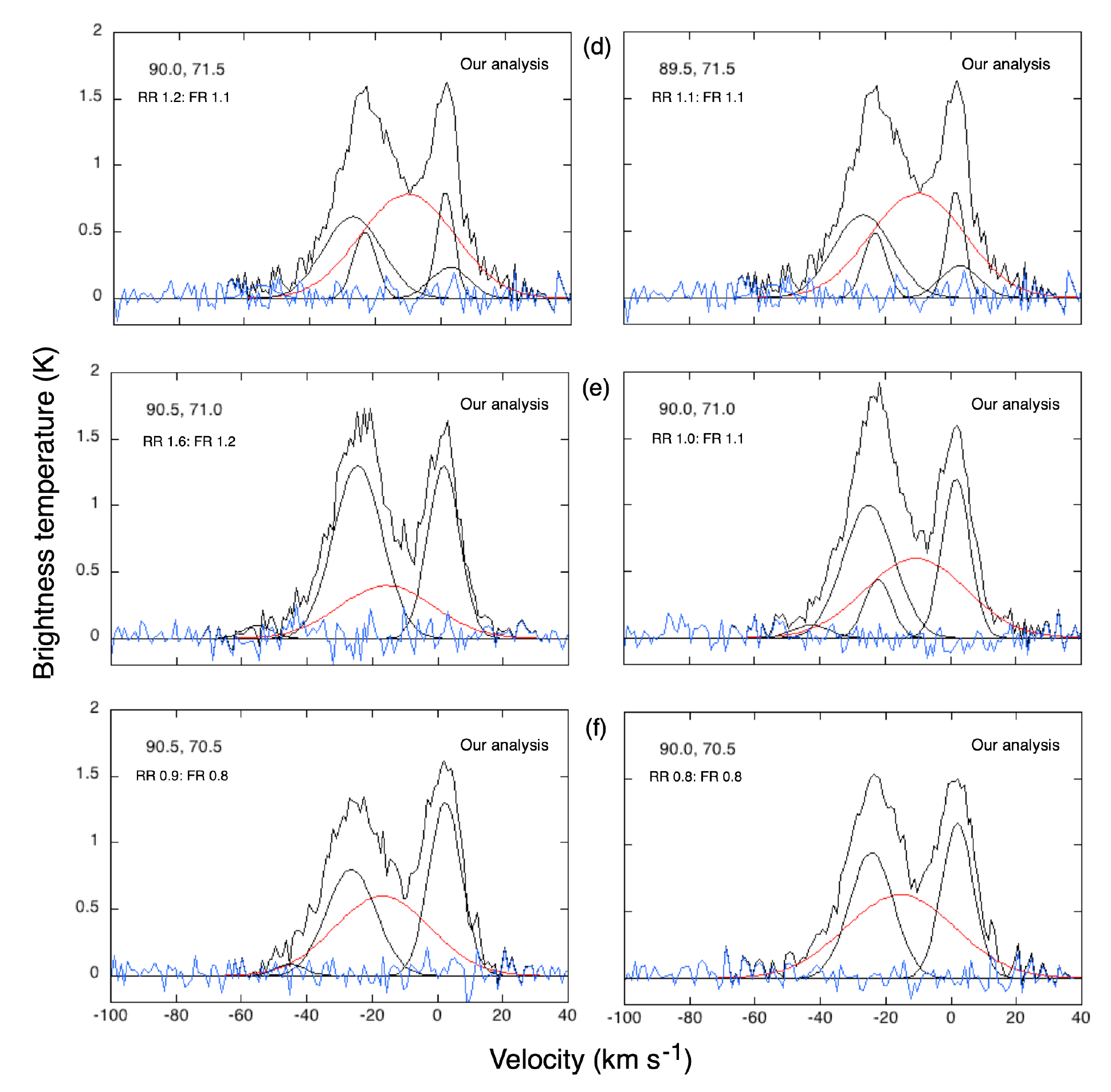}
\caption{The results of our Gauss fit solutions for the same set for closely spaced directions shown in Fig. 3a-c. These plots show a high degree or continuity between adjacent positions which is expected for closely spaced directions on the sky.}
\end{figure}
\clearpage

\begin{figure}
\figurenum{4a-c}
\epsscale{0.9}
\plotone{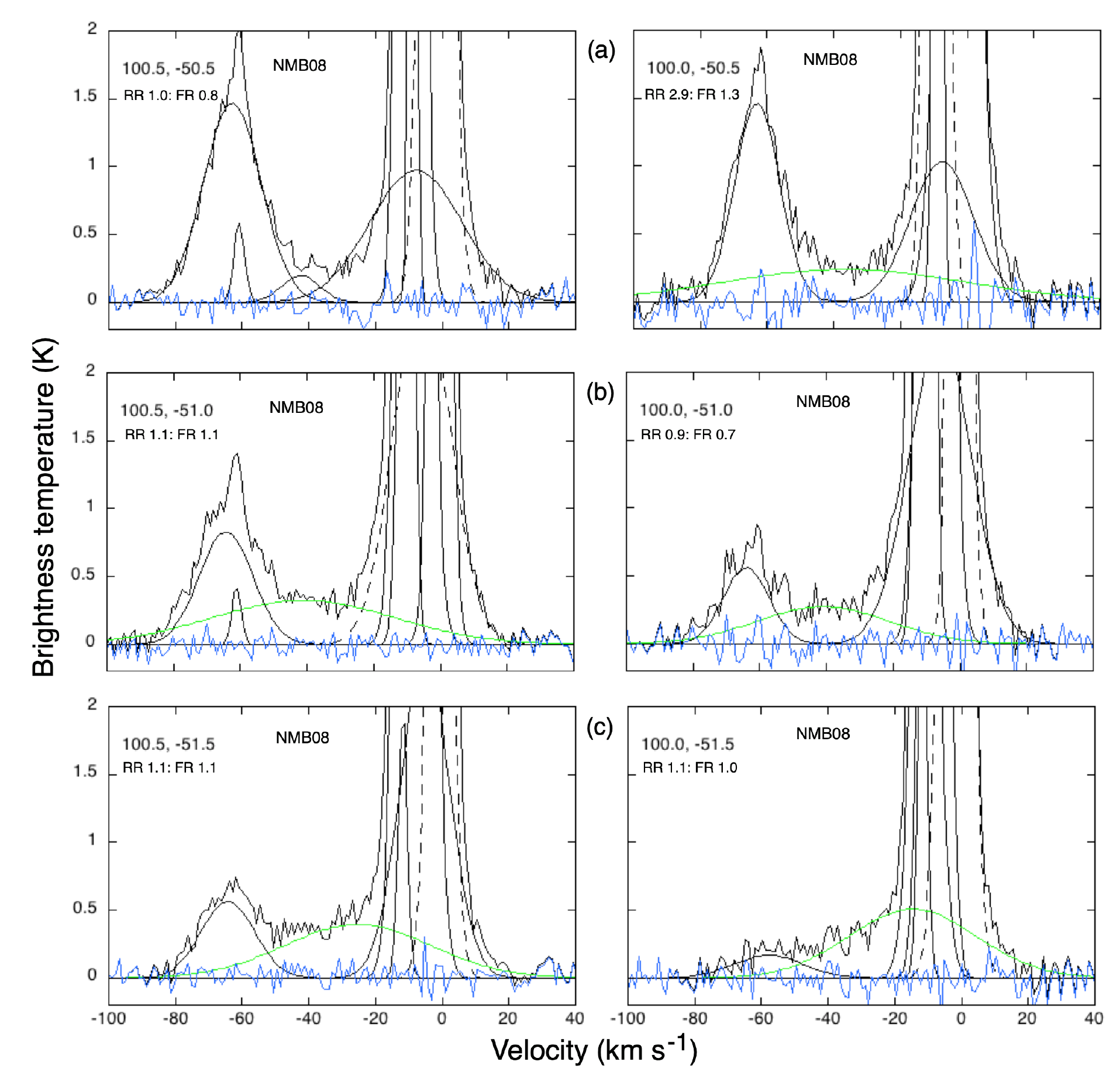}
\caption{Similar to Fig. 3a-c showing the results of the NMB08 analysis for closely spaced directions separated by 0.\arcdeg5 in longitude, which corresponds to half a beam width at these galactic latitudes. As for Figs. 3a-c, adjacent solutions are inconsistent with one another. The ultra-broad components are shown in green.
}
\end{figure}
\clearpage

\begin{figure}
\figurenum{4d-f}
\epsscale{0.9}
\plotone{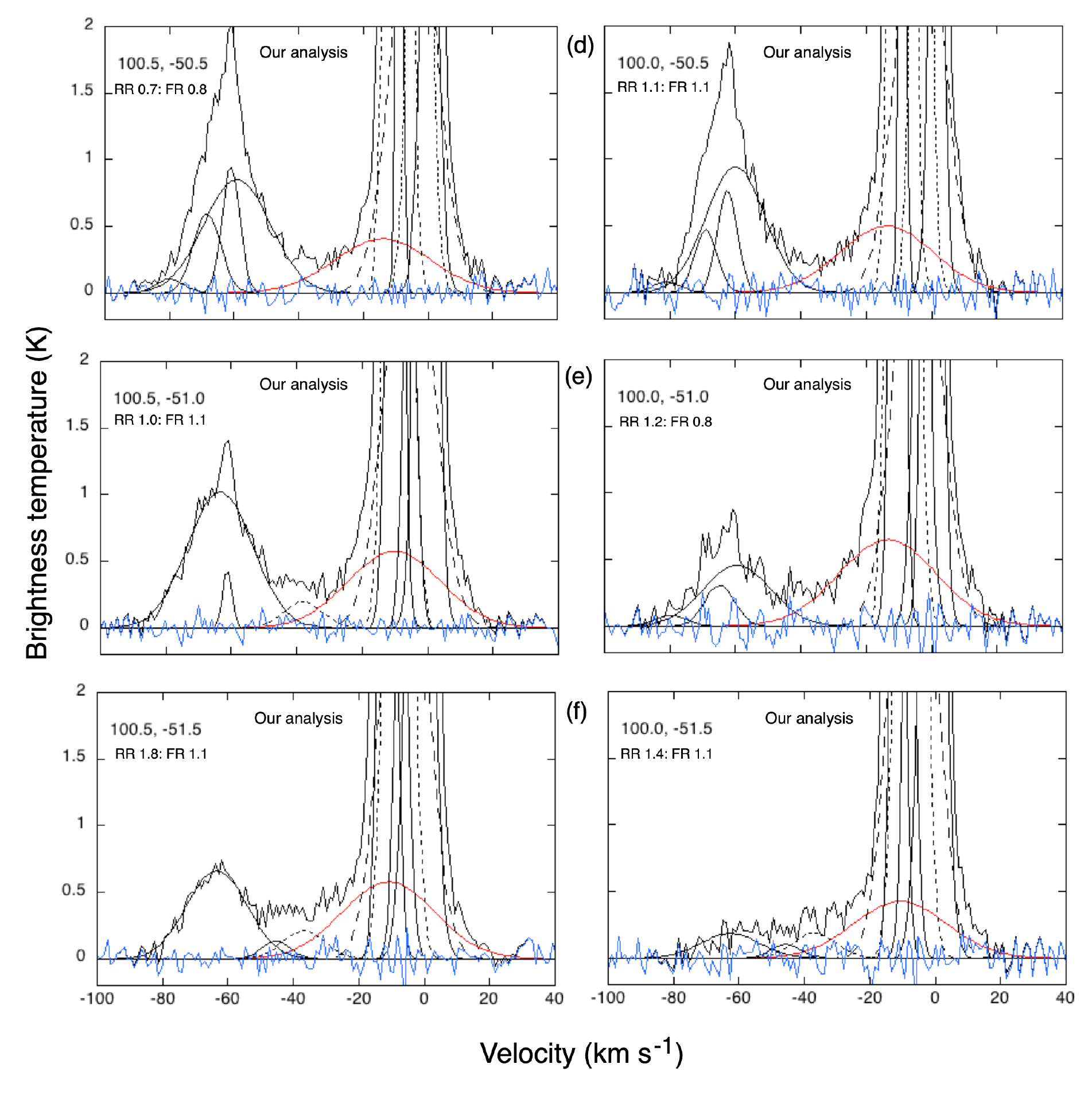}
\caption{Similar to Fig. 3d-f showing our Gauss fit results for the closely spaced directions offset by 0.\arcdeg5 in longitude, which corresponds to half a beam width at these galactic latitudes. Adjacent solutions are again consistent with one another although there are clear differences in pairs of profiles separated by half-a-beamwidth, (e) \& (f) as is evident in the different profiles shapes for some of the pairs. 
}
\end{figure}
\clearpage

\begin{figure}
\figurenum{5}
\epsscale{0.9}
\plotone{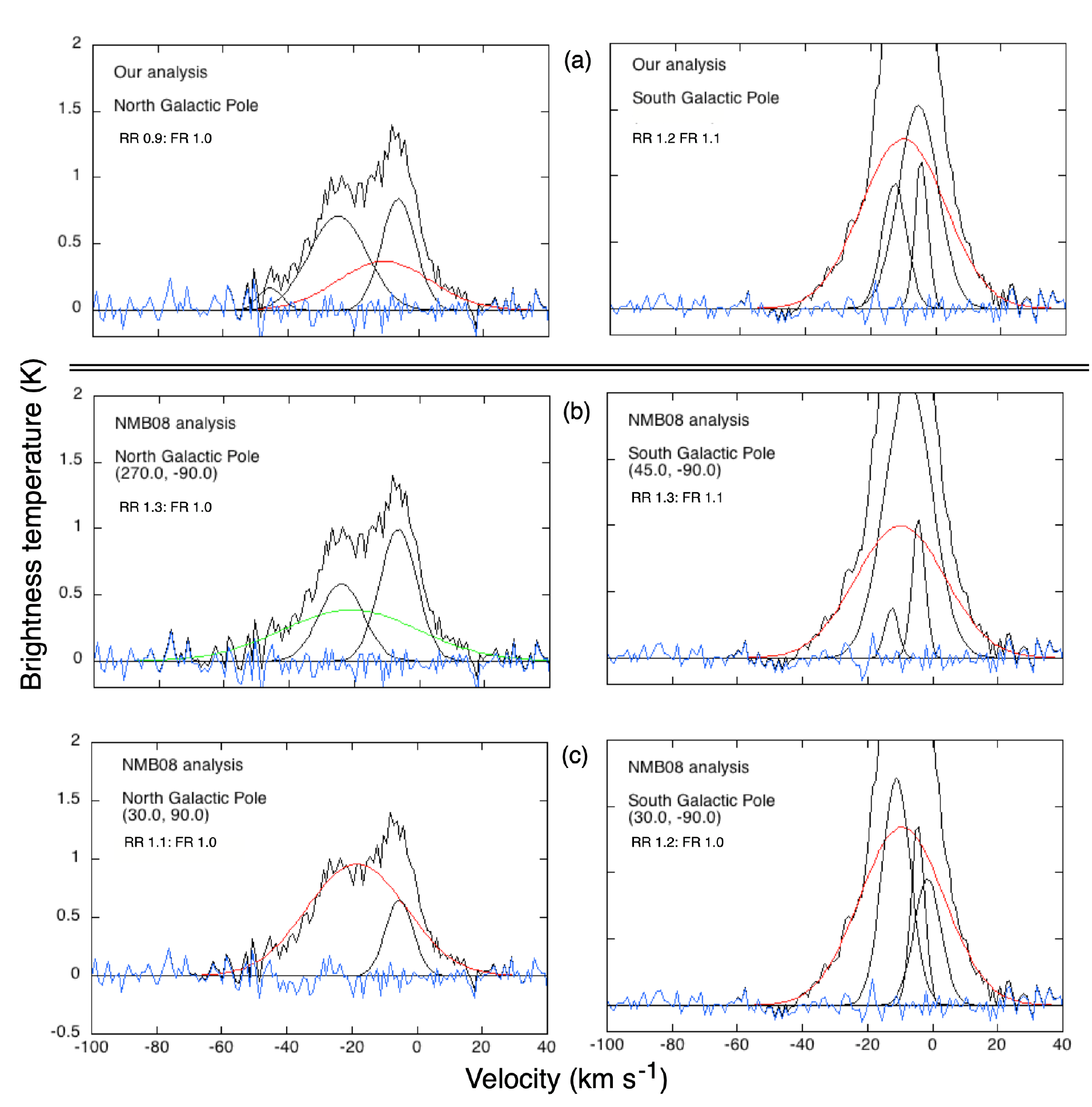}
\caption{Gaussian fit solutions for the North and South Galactic Poles. (a) our results for the North Galactic Pole at the left and for the South Galactic Pole at the right; (b) and (c) show multiple solutions found in the NMB08 database, with longitude and latitude indicated in the plots. The broad components are shown in red, ultra-broad in green, and residuals in blue. Note the discrepancies between the NMB08 solutions for the same direction. 
}
\end{figure}
\clearpage

\begin{figure}
\figurenum{6}
\epsscale{0.9}
\plotone{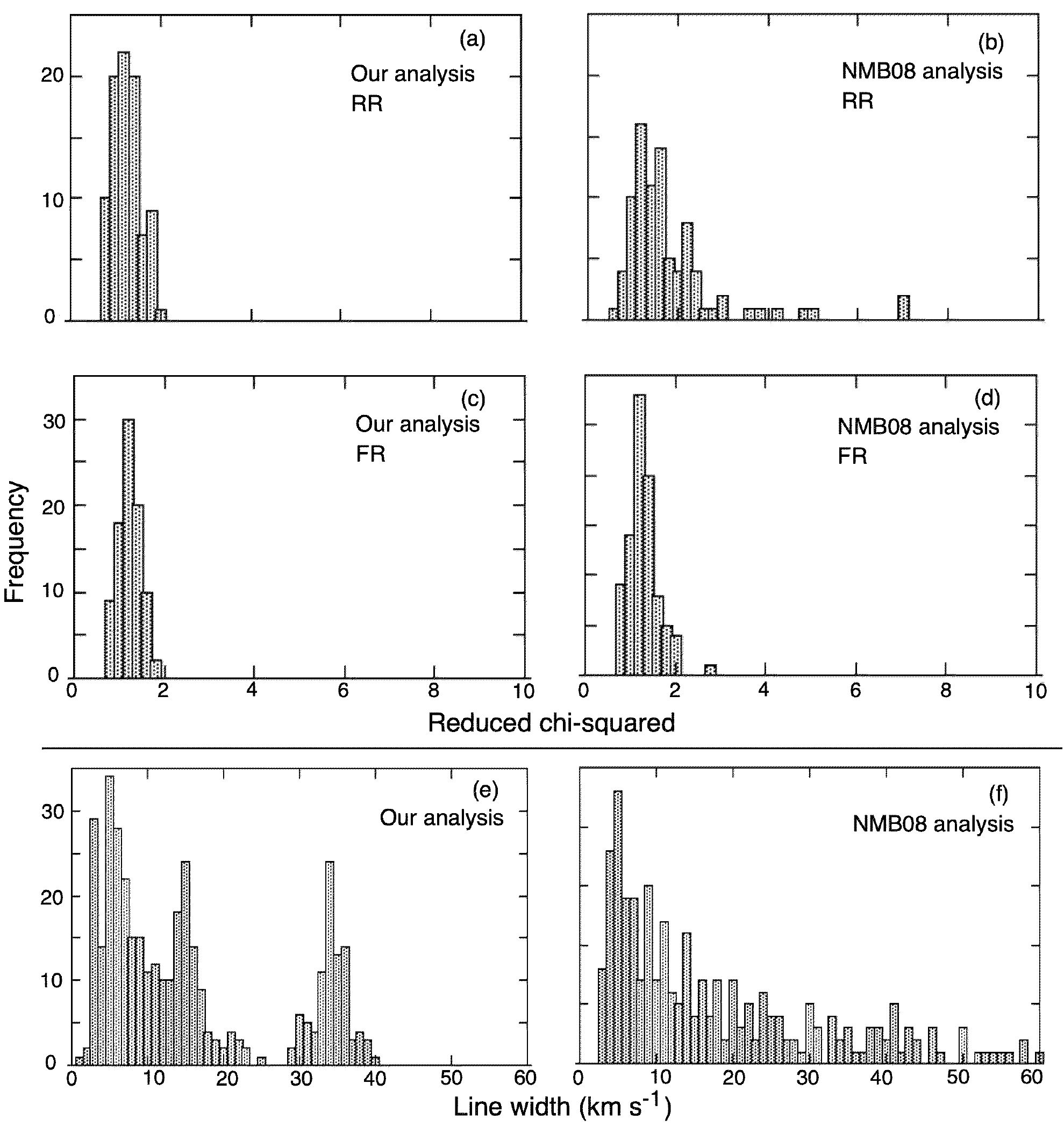}
\caption{Histograms contrasting the results for the two different Gauss fitting methods. The top row shows $\tilde{\chi}^{2}$ where the Restricted Range (RR) is used to measure the goodness of fit for (a) our analysis, where the values are clustered between one and two, and (b) NMB08 analysis, where the values are more spread out and a value of 16.6 is off the scale to the right. The middle row is similar, but uses $\tilde{\chi}^{2}$ for the Full Range (FR), where the values for both (c) our analysis and (d) NMB08 analysis clustered between one and two, illustrating how NMB08 is able to get a $\tilde{\chi}^{2}$ near unity while leaving significant residuals (see text). The bottom row shows histograms of the Gaussian line widths for all components with center velocities between -30 and 30 km s$^{-1}$ for both (e) our analysis and (f) NMB08 analysis. The velocity range was chosen to make a direct comparison with Fig. 3b of Nidever et al. (2008), which also shows a long tail of ultra-broad line widths. This is in striking contrast with our results (e) that show a pronounced peak at 34.2 $\pm$ 1.6 km s$^{-1}$.
}
\end{figure}


\begin{thebibliography}{}
\bibitem[Andrae et al. (2010)]{Andrae 2010}
Andrae, R., Schulze-Hartung, T., \& Melchior, P. 2010, arXiv 1012.3754A
\bibitem[Arnal et al. (2000)]{Arnal 2000}
Arnal, E. M., Bajaja, E., Larrarte, J. J., et al. 2000, A\&AS, 142, 35
\bibitem[Bajaja et al. (2005)] {Bajaja 2005}
Bajaja, E., Arnal, E. M., Larrarte, J. J., et al. 2005, A\&A, 440, 767
\bibitem[Bregman (1980]{Bregman 1980}
Bregman, J.N. 1980, \apj, 236, 577
\bibitem[Clark et al. (2014)]{Clark 2014}
Clark, S. E., Peek, J. E. G., \& Putman, M. E. 2014, \apj, 789, 82
\bibitem[Clark et al. (2015)]{Clark 2015}
Clark, S. E., Colin-Hill, J., Peek, J. E. G., et al. 2015, PRL, 115, 241302
\bibitem[Hartmann \& Burton (1997)]{Hartmann 1997}
Hartmann, D., \& Burton, W. B. 1997, {\it Atlas of Galactic Neutral Hydrogen}, (Cambridge, UK: Cambridge University Press)
\bibitem[Haud (2000)]{Haud 2000}
Haud, U. 2000, A\&A, 364, 83 
\bibitem[Haud \& Kalberla (2007)]{Haud 2007}
Haud, U., \& Kalberla, P. M. W. 2007, A\&A, 466, 555
\bibitem[Henshaw et al. (2016)]{Henshaw 2016}
Henshaw, J. D., Longmore, S. N., Kruijssen, J. M. D., et al. 2016, MNRAS, 457, 2675H
\bibitem[Kalberla et al. (2007)]{Kalberla 2007}
Kalberla, P. M. W., Burton, W. B., Hartmann, D., et al. 2005, A\&A, 440, 775 
\bibitem[Kalberla et al. (2016)]{Kalberla 2016}
Kalberla, P. M. W., Kerp, J., \& Haud, U., et al. 2016, \apj, 821, 117
\bibitem[Kuntz \& Danly (1996)]{Kuntz 1996}
Kuntz, K.D., \& Danly, L. 1996, \apj, 457, 703
\bibitem[Marchal et al. (2019)]{Marchal 2019}
Marchal, A., Miville-Desch\'enes, M-A., Orieux, F., et al. 2019, A\&A, 626, A101
\bibitem[Martin et al. (2015)]{Martin 2015}
Martin, P. G., Blagrave, K.P. M., Lockman, F.J., et al. 2015, \apj, 809, 153 
\bibitem[McClure-Griffiths et al. (2009)]{McClure-Griffiths 2009}
McClure-Griffiths, N. M., Pisano, D. J., Calabretta, M. R., et al. 2009, \apjs, 181, 398
\bibitem[Miville-Desch\'enes et al. (2017)]{Miville 2017}
Miville-Desch\'enes, M-A., Salom\'e, Q., \& Martin, P. G., et al. 2017, A\&A, 599, A109 
\bibitem[Nidever et al. (2008)]{Nidever 2008}
Nidever, D. L., Majewski, S. R., \& Burton, W. B. 2008, \apj, 679, 432
\bibitem[Peek et al (2011)]{Peek 2011}
Peek, J.E.G., Heiles, C., Douglas, K. A., et al. 2011, \apjs,194, 20 
\bibitem[Peratt \& Verschuur (2000)]{Peratt 2000}
Peratt, A. L., \& Verschuur, G. L. 2000, IEEE Trans. Plasma Sci., 28, 2122
\bibitem[Poppel at. al (1994)]{Poppel1994}
P\"oppel, W. G. L., Marronetti, P., \& Benaglia, P. 1994, A\&A, 287, 601
\bibitem[Schmelz et al. (2010)]{Schmelz 2010}
Schmelz, J. T., Saar, S.H., Nasraoui, K., et al. 2010, \apj, 723, 1180
\bibitem[Schmelz et al. (2013)]{Schmelz 2013}
Schmelz, J. T., Pathak, S. Jenkins, B.S., \& Worley, B.T. 2013, \apj 764, 53
\bibitem[Verschuur (2004)]{Verschuur 2004}
Verschuur, G. L. 2004, AJ, 127, 394
\bibitem[Verschuur \& Schmelz (2010)]{Verschuur 2010}
Verschuur, G. L., \& Schmelz, J. T. 2010, AJ, 139, 2410 
\bibitem[Verschuur et al. (2018)]{Verschuur 2018}
Verschuur, G. L., Schmelz, J. T., \& Asgari-Targi, M. 2018, \apj, 867, 139
\bibitem[Verschuur \& Peratt (1999)]{Verschuur 1999}
Verschuur, G. L., \& Peratt, A. L. 1999, AJ, 118, 1252 
\bibitem[Wakker (2001)]{Wakker 2001}
Wakker, B.P. 2001, \apjs, 136, 463
\bibitem[Winkel (2016)]{Winkel 2016}
Winkel,B., Kerp, J., Fl\'oer, L., et al. 2016, A\&A 585, A41 
\end{thebibliography}
\end{document}